\documentclass[twocolumn,prb,superscriptaddress,amsmath,amssymb,aps]{revtex4}
\usepackage{graphicx}
\usepackage{dcolumn}
\usepackage{bm}

\usepackage[utf8]{inputenc}
\usepackage[T1]{fontenc}
\usepackage{mathptmx}
\usepackage{etoolbox}

\usepackage{braket}
\usepackage{xcolor}
\usepackage{subcaption}
\usepackage{caption}
\makeatletter
\def\BState{\State\hskip-\ALG@thistlm}
\makeatother

\begin{document}

\title{Parameterization protocol and refinement strategies for accurate and transferable analytic bond-order potentials: Application to Re}

\author{Aparna P. A. Subramanyam}
\affiliation{Interdisciplinary Centre for Advanced Materials Simulation (ICAMS), Ruhr-Universit{\"a}t Bochum, 44801 Germany}
\affiliation{Theoretical Division, Los Alamos National Laboratory, USA}
\affiliation{Equal contributor}
 
\author{Jan Jenke}
\affiliation{Interdisciplinary Centre for Advanced Materials Simulation (ICAMS), Ruhr-Universit{\"a}t Bochum, 44801 Germany}
\affiliation{Equal contributor}

\author{Alvin N. Ladines}
\affiliation{Interdisciplinary Centre for Advanced Materials Simulation (ICAMS), Ruhr-Universit{\"a}t Bochum, 44801 Germany}

\author{Ralf Drautz}
\affiliation{Interdisciplinary Centre for Advanced Materials Simulation (ICAMS), Ruhr-Universit{\"a}t Bochum, 44801 Germany}

\author{Thomas Hammerschmidt}
\affiliation{Interdisciplinary Centre for Advanced Materials Simulation (ICAMS), Ruhr-Universit{\"a}t Bochum, 44801 Germany}

\begin{abstract}
Interatomic potentials provide a means to simulate extended length and time scales that are outside the reach of ab initio calculations. The development of an interatomic potential for a particular material requires the optimization of the parameters of the functional form of the potential. We present a parameterization protocol for analytic bond-order potentials (BOP) that provide a physically transparent and computationally efficient description of the interatomic interaction. The parameterization protocol of the BOP follows the derivation of the BOP along the coarse-graining of the electronic structure from density-functional theory (DFT) to the tight-binding (TB) bond model to analytic BOPs. In particular, it starts from TB parameters that are obtained by downfolding DFT eigenstates of two-atomic molecules to an $sd$-valent minimal basis. This $sd$-valent Hamiltonian is combined with a pairwise repulsion to obtain an initial binding energy relation. The $s$ electrons are then removed from the Hamiltonian and instead represented by an isotropic embedding term. In the final step, the parameters of the remaining $d$-$d$ interaction, the pair repulsion and the embedding term are optimized simultaneously. We demonstrate that the application of this parameterization protocol leads to a basic BOP for Re with good transferability. We discuss different strategies to refine the basic BOP towards global transferability or towards local accuracy. We demonstrate that homogeneous samplings of the structural phase-space in a map of local atomic environments can be used to systematically increase the global transferability. We also demonstrate the influence of training data-weighting on local accuracy refinements with a Pareto-front analysis and suggest further requirements to select a final BOP. The local accuracy and global transferability of the final BOP is also shown and compared to DFT. \end{abstract}

\maketitle

\section{\label{sec:introduction}Introduction}

Quantum-mechanical calculations on the basis of density-functional theory (DFT) allow computational materials scientists in principle to predict all properties of structural and functional materials. In practice, however, the computational cost of DFT calculations limits this approach typically to system sizes of only a few hundred atoms. 
This limitation puts many material phenomena out of reach for DFT calculations, e.g. plastic deformation, melting, phase transitions. One of the central goals of computational materials science at the atomistic scale is therefore to replace DFT calculations partly with computationally more efficient approaches. The two main approaches are to represent the DFT potential energy surface by numerical interpolation of a large number of data points and by a physical model of the interatomic interaction. 

The physical models require orders of magnitude less parameters to be adjusted in their construction. This leads to a significantly lower demand on the training data but also to less flexibility with regards to optimization to a set of training data. The robustness with regard to the predictions therefore depends on the physical ground of the functional form, the training data, and the parameterization strategy. Their subtle interplay is the core of the challenge of developing reliable parameterizations of physical models of the interatomic interactions~\cite{0965-0393-15-3-008, doi:10.1021/ct5001044, DUFF2015439, Barrett2016, doi:10.1021/acs.jctc.5b00673, 0965-0393-25-5-055003, Ladines-20}.

The different physical models are formulated either as an explicit function of atomic positions for specific types of interactions, e.g., metals~\cite{Finnis-84, PhysRevB.29.6443}, semiconductors~\cite{Stillinger-85, Tersoff-86}, or as more flexible coarse-grained electronic-structure methods like tight-binding (TB)~\cite{Pettifor-76, Sutton-88, PhysRevB.51.12947, PhysRevB.58.7260} or bond-order potentials (BOP)~\cite{Pettifor-89, Horsfield-96, Hammerschmidt-09-IJMR, Drautz-15}. Many published models exhibit quantitative and often qualitative differences in their prediction for the same material, see e.g. Refs.~\onlinecite{Moeller-18, Lysogorskiy-19, Starikov-21} for extensive comparisons and benchmarks. Such comparative assessments are often challenged by the different and not always fully transparent choices of training data and parameterization strategy. 

In this work, we propose a parameterization protocol for a basic model and different strategies towards refined models for the case of coarse-grained electronic-structure methods. We focus on analytic BOPs~\cite{Drautz-06} that are derived by a second-order expansion of the energy functional from DFT to TB to BOP~\cite{Drautz-15} and that have been shown to provide a robust description of transition metals with bcc~\cite{Mrovec-04, Mrovec-07-2, Madsen-11, Mrovec-11, 0953-8984-23-27-276004, 0965-0393-22-3-034002, Cak-14, Ford-14, Egorov-submitted} and fcc/hcp~\cite{Girshick-98-1, Znam-03, PhysRevB.73.064104, Ferrari-19, Katnagallu-19} ground-state structures. 

The parameterization is carried out along an analogous coarse-graining route from downfolding DFT eigenstates of dimers to a TB Hamiltonian~\cite{PhysRevMaterials.5.023801}, the addition of a simple pairwise repulsion term, the replacement of $s$-electrons in the Hamiltonian by a simple embedding function and the approximate solution of the TB Hamiltonian with analytic BOP. A new approach is introduced to analyze the transferability to other crystal structures in a transparent and intuitive way with a recently established map of local atomic environments~\cite{Jenke-18}. Different strategies are compared to refine the resulting basic model obtained with few training data towards global transferability by adding a homogeneous coverage of phase space in the map or towards local accuracy by adding specific training data motivated by potential applications.

For the purpose of demonstrating our parameterization protocol, we choose Re with hcp as the ground state as example material in this work. Re is often added to Ni-based superalloys to improve their creep properties \cite{RAE20014113}. In fact, the aircraft industry accounts for almost 70\% of the world's consumption of industrially produced Re \cite{Rhythms_Re}. It also plays a role as product of nuclear transmutation of W under neutron irradiation of plasma-facing materials for divertors in fusion reactors \cite{Gilbert_2011}. In both applications, Re promotes the formation of complex intermetallic phases, particularly topologically close-packed (TCP) phases \cite{doi:10.1021/ja1091672, HASEGAWA20141568, FUKUDA2014460, LLOYD2022101370, RAE20014113}, that deteriorate the mechanical properties. There are only two existing physical models for Re based on EAM~\cite{doi:10.1063/1.4982361, doi:10.1063/1.5030113}. 

In Ref. \onlinecite{doi:10.1063/1.4982361}, the model is fitted mainly to the elastic constants, the equation of state data for the ground state hcp structure along with the cohesive energies and lattice parameters of bcc, fcc and hcp crystal structures. The resulting model describes the elastic constants well, but the self-interstitials were poorly described as shown in Ref. \onlinecite{doi:10.1063/1.5030113}. In Ref. \onlinecite{doi:10.1063/1.5030113}, the Re model is explicitly fitted to several defect, liquid and other crystal structures. This model performs very well for point defects but poorly describes the elastic constants with differences to reference experimental values reaching as high as 270 GPa for some elastic constants. In this regard, we show that our model for Re balances all the properties of interest satisfactorily.

A brief summary of methodology and reference data is given in Sec.~\ref{sec:methodology}. In Sec.~\ref{sec:protocol}, we outline the parameterization along the same coarse-graining route by downfolding DFT eigenstates of dimers to a TB Hamiltonian which is then simplified and solved approximately with analytic BOP. In Sec.~\ref{sec:refinement}, the resulting basic BOP is refined with different strategies towards global transferability and towards local accuracy. The compromise between the two strategies is worked out by identifying the Pareto front for different weighting of training data and by determining the transferability across a broad range of local atomic environments. Using additional tests, a final BOP for Re is selected in Sec.~\ref{sec:selection} and compared to DFT reference data. 

\section{\label{sec:methodology}Methodology}

\subsection{\label{sec:BOP} Analytic bond-order potentials}

The analytic bond-order potentials are derived by coarse-graining the description of the electronic structure from DFT to the tight-binding (TB) bond model~\cite{Sutton-88} to BOP~\cite{Drautz-06,Drautz-15}. For a non-magnetic, charge-neutral system, the most basic form of the total binding energy ($E_\mathrm{bind}$) in the TB bond model is written as 
\begin{equation}
E_\mathrm{bind} = E_{\mathrm{bond}} + E_{\mathrm{rep}}
\end{equation}
where $E_{\mathrm{bond}}$ and $E_{\mathrm{rep}}$ are bonding and repulsive energy, respectively. Further terms due to magnetism and charge transfer~\cite{Drautz-15,Hammerschmidt-19} are not required for the application to Re in this work.

The bond energy $E_{\mathrm{bond}}$ is calculated by integrating the local density of states $n_{i\alpha}(E)$, of orbital $\alpha$ on atom $i$ up to the Fermi level $E_F$ as 
\begin{equation}
\label{eq:Ubond}
E_{\mathrm{bond}} = \sum_{i\alpha} \int_{-\infty}^{E_{F}}(E-E_{i\alpha})n_{i\alpha}(E)dE
\end{equation}
with the energy of the atomic onsite level $E_{i\alpha}$. In the BOP formalism, $n_{i\alpha}(E)$ is not obtained by diagonalization of the TB Hamiltonian $\hat{H}$ but rather by using the moments theorem~\cite{CyrotLackmann-67} that relates the local electronic structure in terms of $n_{i\alpha}(E)$ to the local atomic structure. The $N^{th}$ moment of $n_{i\alpha}(E)$ given by
\begin{equation}
\mu_{i\alpha}^{(N)} = \int E^N n_{i\alpha}(E) dE,
\end{equation}
can also be written as product of pairwise Hamiltionian matrices along self-returning paths that start and end on atom $i$ orbital $\alpha$
\begin{equation}\label{eq:moments}
\begin{split}
\mu_{i\alpha}^{(N)} & = \braket{i\alpha|\hat{H}^{N}|i\alpha}\\
 & = \sum_{j\beta \cdots}
\braket{i\alpha|\hat{H}|j\beta}\braket{j\beta|\hat{H}|k\gamma}\braket{k\gamma|\hat{H}|\cdots}\braket{\cdots|\hat{H}|i\alpha} \\
& = H_{i\alpha j\beta}H_{j\beta k\gamma}H_{k\gamma \cdots}H_{\cdots i\alpha}
\end{split}
\end{equation}
where $H_{i\alpha j\beta}$ is the TB Hamiltonian that describes the interaction between atom $i$ orbital $\alpha$ and atom $j$ orbital $\beta$. Details of the computation of $E_{\mathrm{bond}}$ from $H_{i\alpha j\beta}$ are given elsewhere~\cite{Hammerschmidt-19}. The construction of $H_{i\alpha j\beta}$ is a central part of the TB/BOP parameterization. For most BOP models, the values of $H_{i\alpha j\beta}$ are derived from DFT calculations and kept fixed in the parameterization~\cite{PhysRevB.73.064104, Cak-14, 0965-0393-22-3-034002}. The TB Hamiltonians for $sd$-valent systems are often simplified by replacing the $s$ contribution in the Hamiltonian with an additional attractive embedding energy $E_{\mathrm{emb}}$~\cite{PhysRevB.83.184119, 0953-8984-23-27-276004}, as described in detail below.

The attractive bond energy of a TB/BOP model is balanced by repulsive contributions $E_{\mathrm{rep}}$ that represent the overlap repulsion of atomic orbitals and higher order terms. The simplest form of $E_{\mathrm{rep}}$ is a repulsive pair potential. More sophisticated forms of $E_{\mathrm{rep}}$ like a Yukawa-like term~\cite{Uenv_Nguyen_Manh, 0953-8984-19-23-236228} account for environment-dependent many-body repulsion~\cite{PhysRevB.73.064104, 0965-0393-22-3-034002}. The interaction range of the TB/BOP model is limited by multiplying bond integrals and pair repulsion with a cut-off function 
\begin{equation}
 f_\mathrm{cut} = \frac{1}{2}\left[ \cos\left(\pi \frac{R_{ij}-(r_\mathrm{cut}-d_\mathrm{cut})}{d_\mathrm{cut}}\right) + 1 \right]
\end{equation}
in the range of $[r_\mathrm{cut} - d_\mathrm{cut}$, $r_\mathrm{cut}]$. For Re in this work, we use $r_{\mathrm{cut}}=6$ \AA~ and $d_{\mathrm{cut}}=0.5$ \AA~. 

The further settings of the analytic BOP model developed in this work follow previous parameterizations for transition metals~\cite{Cak-14, Ford-14, Katnagallu-19, Ferrari-19, Egorov-submitted}.
In particular we use 9 calculated moments, higher moments estimated up to 100 and a square-root terminator with Jackson kernel to ensure a strictly positive DOS. The calculations are performed self-consistently with numerically enforced charge-neutrality. The calculations are carried out with the \texttt{BOPfox} software \cite{Hammerschmidt-19}.

\subsection{Parameterization setup}

The parameterization of an interatomic potential in general or a TB/BOP model in particular, requires to adjust the free parameters such that the chosen set of training data is reproduced with sufficient accuracy. The underlying numerical procedure uses a cost function that measures the discrepancy between the predictions of the potential and the reference data. A natural choice for the cost function is the root-mean-square (RMS) error
\begin{equation}
c(\bm{\theta}) 
= \sqrt{
\frac{\sum_{n}^{N_{\mathrm{ref}}} e_n^2}{N_{\mathrm{ref}}}
} ,
\label{eq:cost_function}
\end{equation}
where 
\begin{equation}
e_n = w_n \left( E_{\mathrm{pred}, n} (s_n; \bm{\theta}) - E_{\mathrm{ref}, n}(s_n) \right).
\label{eq:error_cost_function}
\end{equation}
$N_{\mathrm{ref}}$ is the number of reference data points, $\bm{\theta}$ is a vector of model parameters, $E_{\mathrm{pred}, n} (s_n; \bm{\theta})$ is the model prediction for structure $s_n$ for given model parameters. $E_{\mathrm{ref}, n}$ is the corresponding reference and $w_n$ a weight factor to balance the relative importance of the reference data. 
We use a local minimization procedure, the Levenberg-Marquardt algorithm \cite{10.2307/43633451, 10.2307/2098941,osti_7256021}, to minimize the cost function $c(\bm{\theta})$. The minimization is implemented in the \texttt{BOPcat} software~\cite{Ladines-20} that drives BOP calculations with the \texttt{BOPfox} software.
The parameterization of the BOP model is carried out in two steps. In the first step, a basic BOP model is constructed with a small set of training data. In the second step the basic BOP model is refined and validated against the full set of reference data. 

\subsection{\label{sec:referencedata}Reference data}

The reference data for the parameterization of the analytic BOP for Re in this work are total energies obtained from DFT calculations. These energies are computed using non-spin-polarized DFT calculations with \texttt{VASP} \cite{KRESSE199615, PhysRevB.54.11169, PhysRevB.59.1758}. The projector augmented-wave method (PAW) \cite{PhysRevB.50.17953} method is used with the generalized gradient approximation~\cite{PhysRevLett.77.3865}. High accuracy of the calculations is obtained by a plane-wave cut-off energy of 400\,eV and Monkhorst-Pack \cite{PhysRevB.13.5188} k-point meshes with linear density of 0.125\,\AA$^{-1}$. 

\begin{figure}[ht]
\includegraphics[width=0.9\columnwidth]{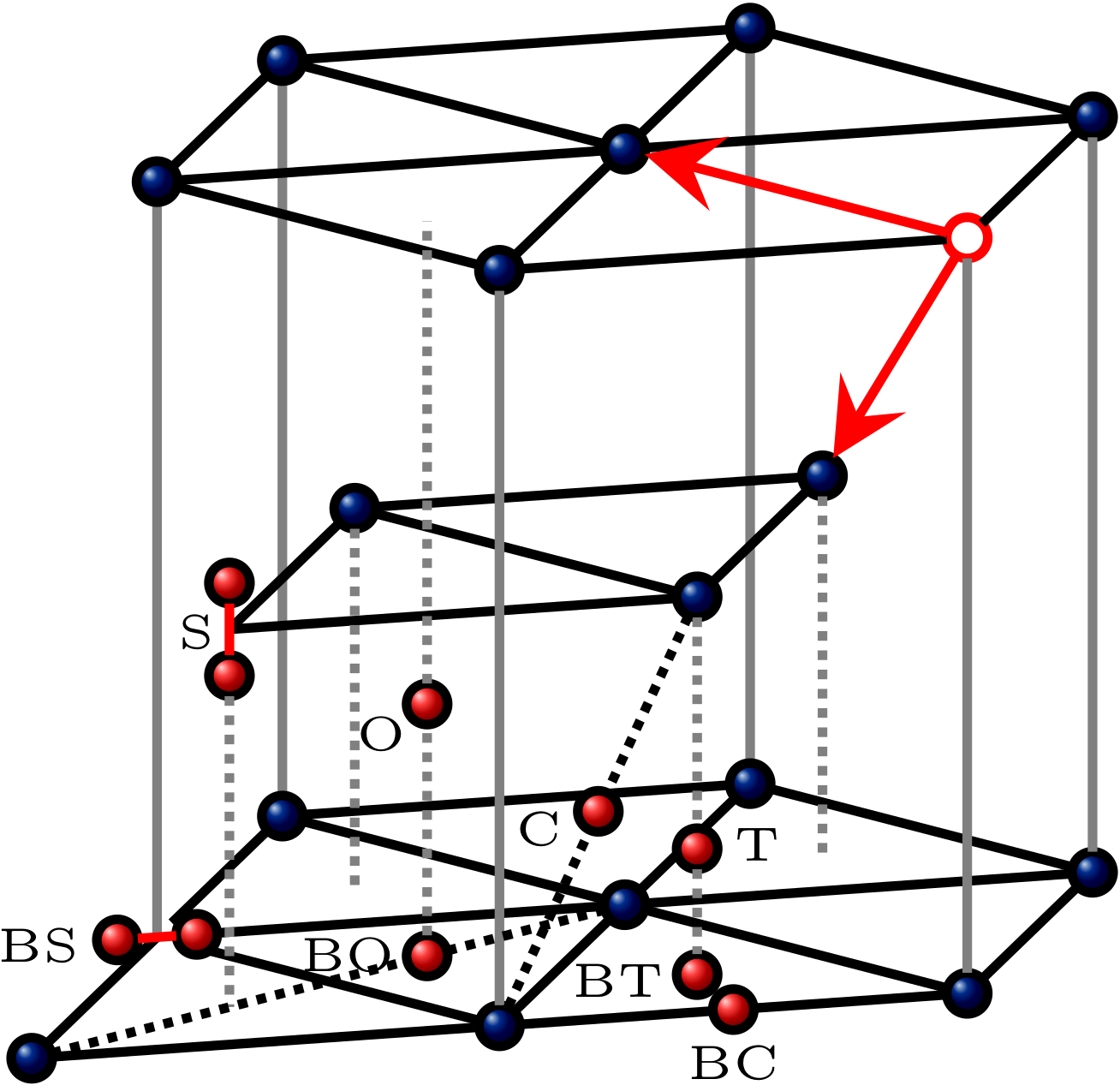}
\caption{Point defects in hcp crystal structures: Self-interstitial atoms in crowdion (C), octahedral (O), split dumbbell (S), tetrahedral (T), basal crowdion (BC), basal octahedral (BO), basal split dumbbell (BS) and basal tetrahedral (BT) configuration (red atoms), vacancy (red circle) and vacancy diffusion-paths (red arrows). Adapted from Ref. \onlinecite{doi:10.1063/1.5030113}.}
\label{fig:hcp_with_defects}
\end{figure}

The reference data covers ideal crystal structures including the basic structures hcp, dhcp, fcc, bcc as well as the topologically close-packed (TCP) phases A15, C14, C15, C35, $\sigma$, $\mu$, $\chi$ that are known to form as Re-compounds (see e.g. Refs.~\onlinecite{doi:10.1021/ja1091672,Hammerschmidt-13}) or as Re-containing precipitates in Ni-based superalloys (see e.g. Ref.~\onlinecite{Rae-01}).
The energy-volume curves for the different crystal structures are computed for 20 structures within $\pm$20\% of the  equilibrium volume and fitted to fifth-order polynomials for obtaining the equilibrium volume, energy and bulk modulus. For the hcp ground state, we additionally include the elastic constants (using 14 structures within strain rates of up to $\pm$2\% along each elastic deformation) and the phonon spectrum. The energy-volume curves for all the crystal structures along with the elastic constants for the ground state hcp structure were obtained using pyiron~\cite{JANSSEN201924}.
In order to assess the transferability of the BOP model we furthermore consider (i) vacancies and the different self-interstitial atom (SIA) configurations shown in Fig. \ref{fig:hcp_with_defects}, (ii) vacancy diffusion-paths within the basal plane and perpendicular to it computed with the nudged-elastic-band method~\cite{doi:10.1063/1.1329672, doi:10.1063/1.1323224} as well as (iii) the basal intrinsic stacking fault and the extrinsic stacking fault that are related to plastic deformation~\cite{hull2011introduction,HU20131136,YIN2017223}.

\section{\label{sec:protocol}Parameterization of the initial BOP model}

\subsection{Initial guess}

The BOP model, as many other interatomic potentials, corresponds to a non-linear relation between atomic structures and their energy. One can therefore expect multiple local minima of the cost function (Eq.~\ref{eq:cost_function}) in the space of model parameters. Searching the global minimum usually involves prohibitive computational cost and therefore local minimization algorithms are common practice. This leads, however, to a potential dependence of the optimized model parameters on the initial guess of their values that is needed to start the local minimization. A physically sound initial guess is therefore required as it more likely leads to a physically meaningful local minimum.

For BOP models, we can construct a physically motivated initial guess for the Hamiltonian $H_{i\alpha j\beta}$ by utilizing downfolded DFT eigenstates of Re-Re dimers to a minimal basis with $sd$ orbitals~\cite{PhysRevMaterials.5.023801}. 
The $ss\sigma$, $sd\sigma$, $dd\sigma$, $dd\pi$, and $dd\delta$ matrix elements of the orthogonal $sd$-valent Hamiltonian are computed for different bond lengths and parameterized as
\begin{equation}
    H(R) = \sum_i c_i \mathrm{exp}(-\lambda_iR^{n_i})
\label{eq:beta}
\end{equation}
with $R$ the distance between the two atoms of the Re-Re dimer.
The onsite matrix elements are taken from the values of the free atom computed from the asymptotic values of the parameterizations for large dimer bond-lengths $R$.
The downfolded $H_{i\alpha j\beta}$ show good ad hoc transferability to different crystal structures~\cite{PhysRevMaterials.5.023801, PhysRevB.83.184119, 0953-8984-23-27-276004, Katnagallu-19, Egorov-submitted} and will be further optimized during the parameterization process in this work.
This work focuses on Re but the approach is general and the downfolded $H_{i\alpha j\beta}$ are available in a database for all homovalent and heterovalent dimers across the periodic table~\cite{PhysRevMaterials.5.023801}.

Further parameters to be set are the number of $s$ and $d$ valence electrons of the BOP model. We estimate the number of electrons with $s$ and $d$ character by projecting the DOS obtained from DFT calculations for hcp-Re on $s$ and $d$ orbitals, respectively. The resulting values are 0.77 $s$-electrons and 5.3 $d$-electrons. During the optimization procedure we adjust the number of $d$-electrons to 5.7. The total number of valence electrons in the BOP model  $N_{\mathrm{e}} = 6.47$ is in close agreement with the $5d^56s^2$ electronic configuration of the Re pseudo-potential of the DFT calculations.

\subsection{Parameterization protocol}

\begin{figure}
\includegraphics[angle=270,width=\columnwidth]{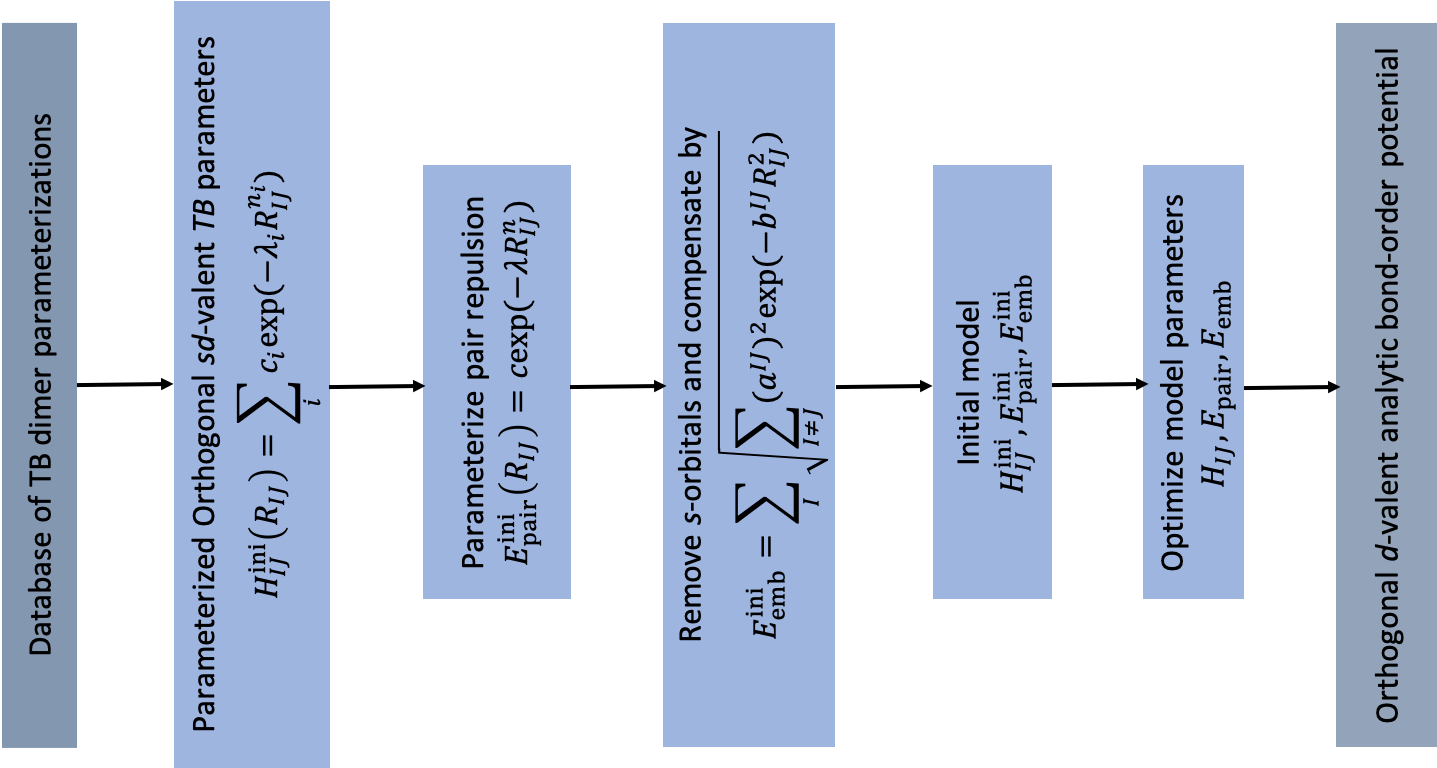}
\caption{Parameterization protocol for constructing a basic BOP model for $sd$-valent systems.}
\label{fig:parameterization_strategy}
\end{figure}

The BOP methodology is a coarse-grained description of the electronic structure and thereby provides a certain degree of intrinsic robustness and transferability unlike simple interatomic potentials on one side and data-based machine-learning potentials on the other side. The goal of the parameterization protocol in this section is to construct a basic BOP parameterization with minimal model complexity and minimal training data. Due to the underlying physics, the basic BOP reaches good robustness and transferability already at this low level of parameterization. From this level, the basic BOP can be further refined by increasing model complexity and training data as shown in the next section or serve as common starting point for the development of compound models.
The parameterization protocol is summarized in Fig.~\ref{fig:parameterization_strategy} and explained step-by-step in the following. The only training data used for the basic BOP are the energy-volume data of hcp, fcc and bcc crystal structures. The parameterization progress in describing this minimal set of training data during execution of the parameterization protocol is compiled in Fig.~\ref{fig:parameterization_procedure}.

\subsubsection{Step 1: Initial sd-valent Hamiltonian}

\begin{figure}[ht]
\begin{subfigure}[b]{0.48\columnwidth}
\includegraphics[width=\textwidth]{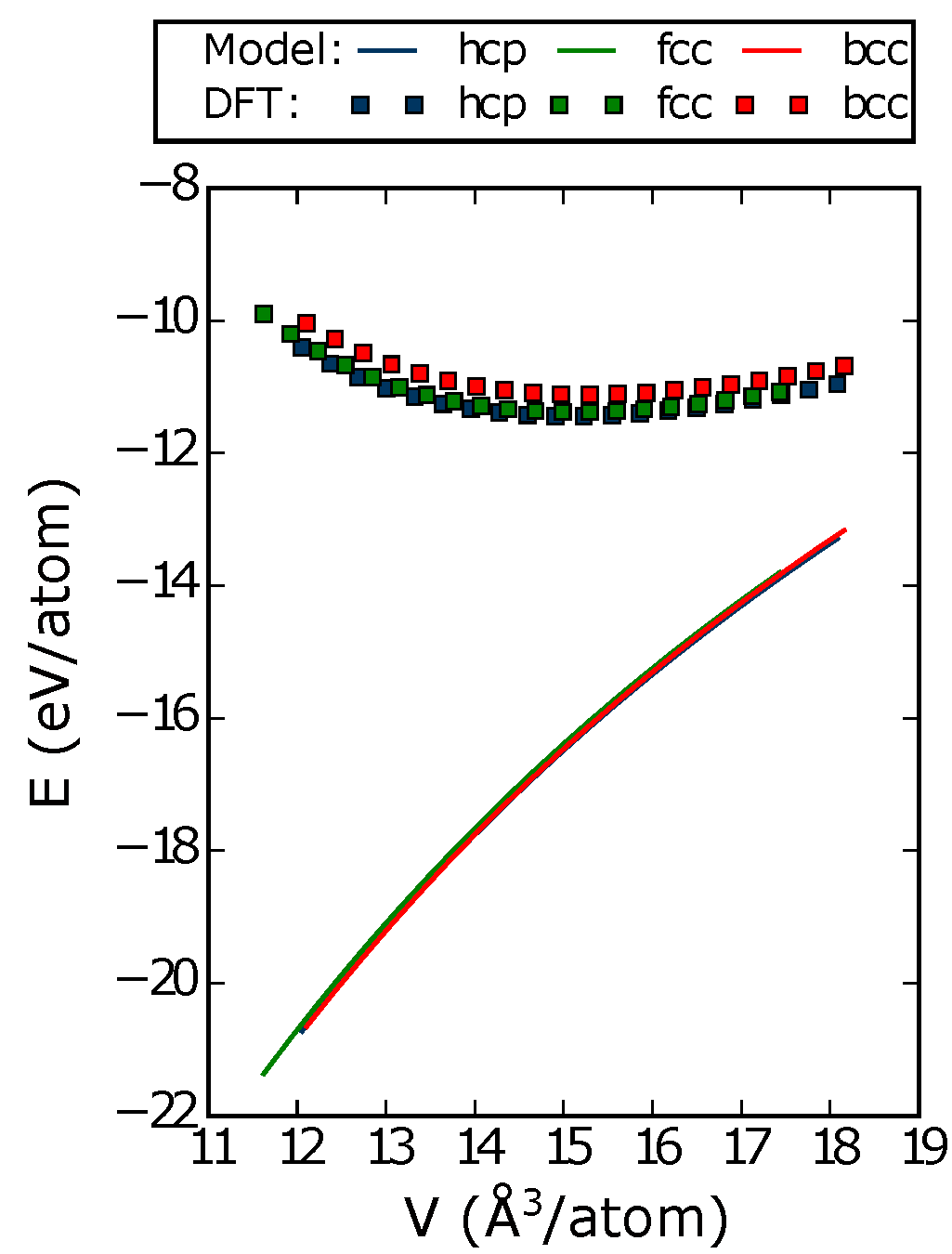}
\caption{a) $H_{IJ}^{\mathrm{ini}, sd}$}
\label{fig:Hini}
\end{subfigure}
\begin{subfigure}[b]{0.48\columnwidth}
\includegraphics[width=\textwidth]{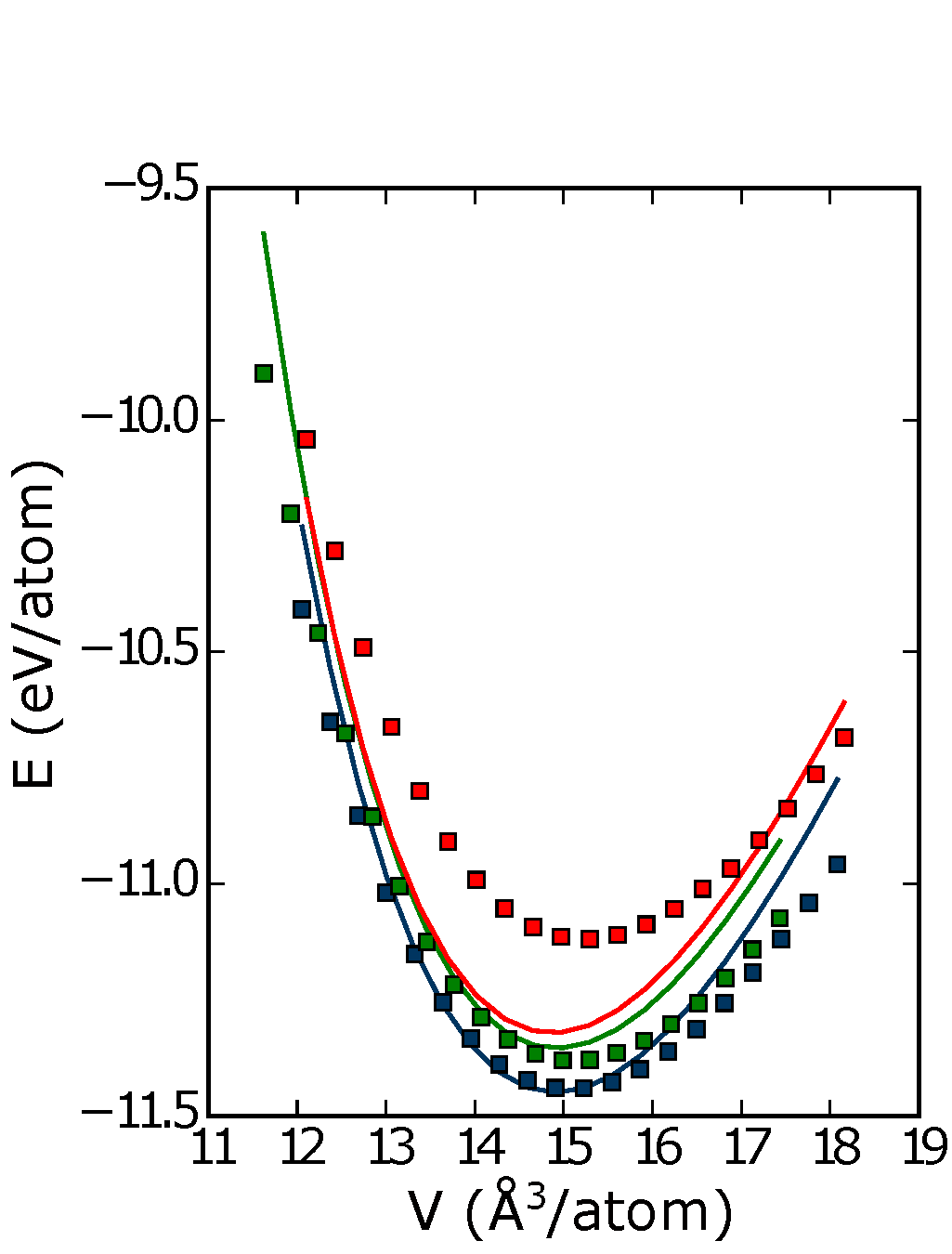}
\caption{b) $H_{IJ}^{\mathrm{ini}, sd}$, $E_{\mathrm{rep}}^{\mathrm{ini}}$}
\label{fig:H_Urep_ini}
\end{subfigure}
\begin{subfigure}[b]{0.48\columnwidth}
\includegraphics[width=\textwidth]{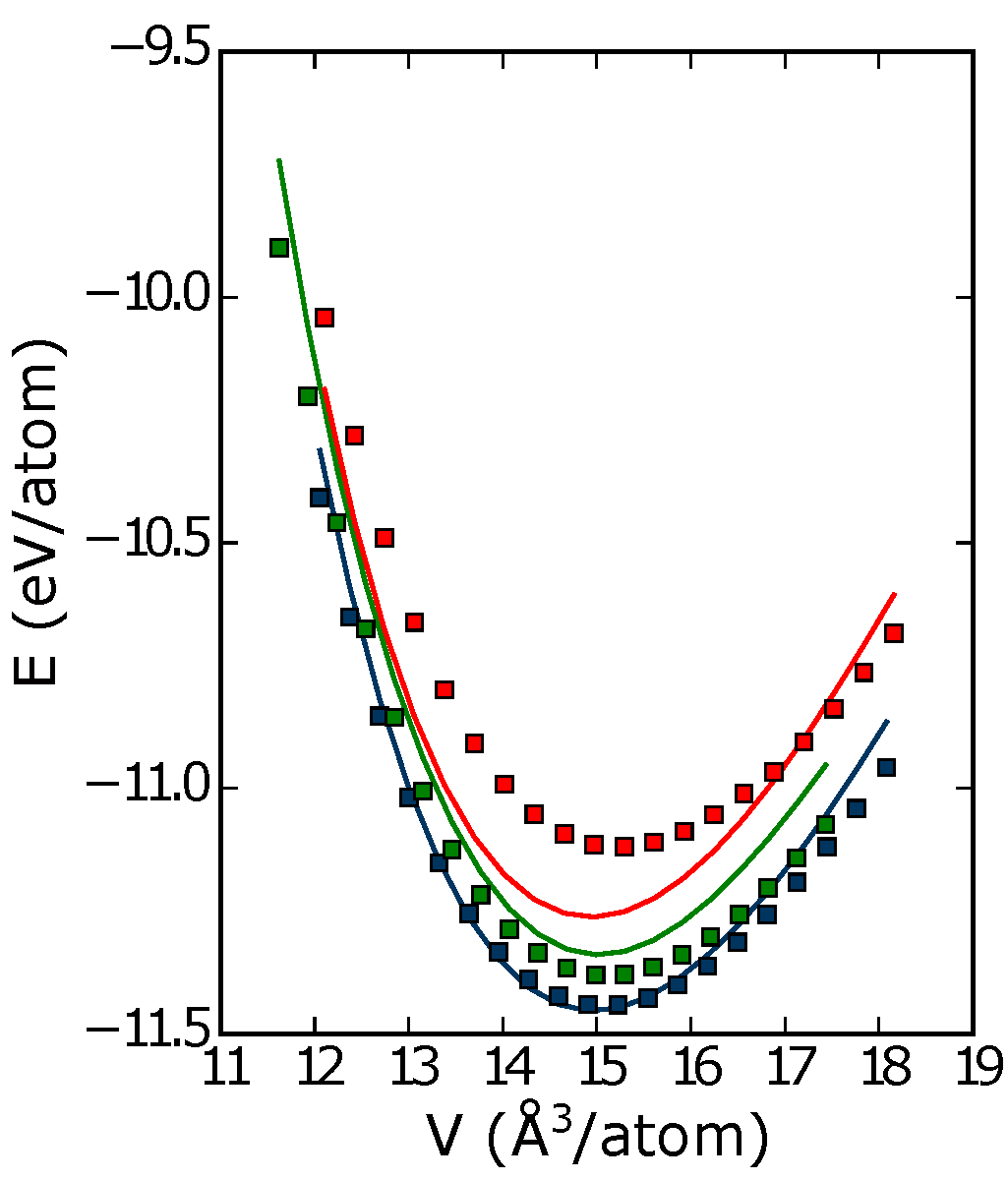}
\caption{c) $H_{IJ}^{\mathrm{ini}, d}$, $E_{\mathrm{rep}}^{\mathrm{ini}}$, $E_{\mathrm{emb}}^{\mathrm{ini}}$}
\label{fig:H_Urep_Uemb_ini}
\end{subfigure}
\begin{subfigure}[b]{0.48\columnwidth}
\includegraphics[width=\textwidth]{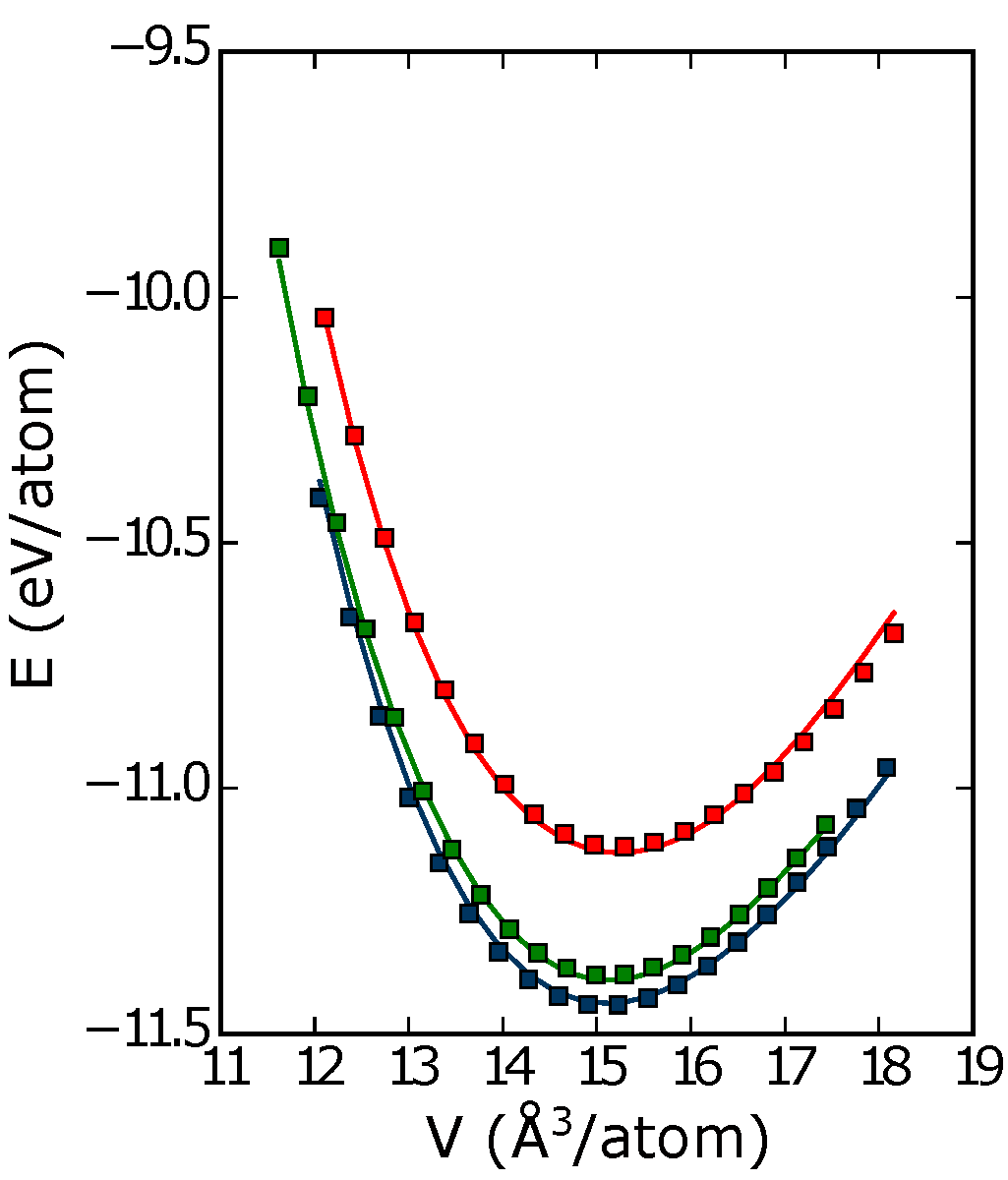}
\caption{d) $H_{IJ}^{d}$, $E_{\mathrm{rep}}$, $E_{\mathrm{emb}}$}
\label{fig:H_Urep_Uemb}
\end{subfigure}
\caption{Parameterization progress of basic BOP model with minimal complexity and minimal reference data. Starting from (a) the initial $sd$ Hamiltonian $H_{IJ}^{\mathrm{ini}, sd}$, (b) the pairwise repulsion $E_{\mathrm{rep}}^{\mathrm{ini}}$ is added and optimized to the DFT reference data, (c) the energy contribution of  $s$-electrons is replaced by an embedding term $E_{\mathrm{emb}}^{\mathrm{ini}}$ and (d) all parameters are optimized to the basic BOP model with $H_{IJ}^{d}$, $E_{\mathrm{rep}}$ and $E_{\mathrm{emb}}$.}
\label{fig:parameterization_procedure}
\end{figure}

In the first step of the parameterization protocol, the BOP model includes only the bond energy (Eq.~\ref{eq:Ubond}) computed with the initial $sd$-valent Hamiltonian $H_{IJ}^{\mathrm{ini}, sd}$ from the downfolded DFT eigenstates. In the absence of a repulsive counterpart, the energy-volume curves exhibit no minimum and cannot be compared directly to DFT as shown in Fig.~\ref{fig:Hini}. The differences in the BOP and the DFT energy-volume curves are comparable, however, by using the structural energy-difference theorem~\cite{Pettifor-86-1,Pettifor-book}. For the development of BOP models, this is particularly useful for verification of the number of valence electrons (see e.g. Refs.~\onlinecite{Seiser-11,Cak-14}) in an early stage of the parameterization.

\subsubsection{Step 2: Addition of repulsive pair potential}

In the second step, the BOP model is extended to include a repulsive part that counteracts the purely attractive bond energy of the $sd$ Hamiltonian. With the explicit treatment of $s$-electrons in the Hamiltonian, we add only a pairwise term with flexible functional form similar to the Hamiltonian matrix elements (Eq. \ref{eq:beta}) as repulsive part 
\begin{equation}
E_{\mathrm{pair}} = \sum_{I\neq J} c_\mathrm{rep}\exp(-\lambda_{\mathrm{rep}} R_{IJ}^{n_{\mathrm{rep}}})
\label{eq:pair}
\end{equation}
with the distance $R_{ij}$ between atoms $i$ and $j$. The parameters $c_\mathrm{rep}$, $\lambda_{\mathrm{rep}}$ and $n_{\mathrm{rep}}$ are adjusted by optimizing the cost function (Eq.~\ref{eq:cost_function}) for the hcp, fcc and bcc crystal structures while $H_{IJ}^{\mathrm{ini}, sd}$ is kept fixed. In this step and in step 3 we performed optimizations with different random initialization in order to verify that the optimization converges to the same minimum. 

Already at this level, the BOP model is able to capture the overall character of the interatomic interaction. This manifests in the correct qualitative energetic ordering of the three crystal structures (see Fig.~\ref{fig:H_Urep_ini}) and the correct range of formation energies. The quantitative performance will be improved in step 4.

\subsubsection{Step 3: Removal of $s$-orbitals in Hamiltonian}

Step 3 of the parameterization protocol is devoted to a simplification of the $sd$ Hamiltonian by removing the $s$-electrons that do not play a significant role for a transition-metal like Re. Other chemical elements may require an explicit treatment of the $s$-electrons, particularly if the interatomic interaction is governed by $sd$-hybridization. This step is computationally attractive not only with regard to the size of the Hamiltonian, but also allows a lowering of the cutoff if only the less extended $d$ orbitals participate in the interaction that directly translates to an increase in computational efficiency. 
The $s$-electrons are removed from the BOP model by taking out the corresponding $ss$ and $sd$ matrix elements from the Hamiltonian and by adjusting the number of valence electrons to $N_{\mathrm{e}} = 5.7$. The cutoff of the BOP model is reduced from $r_{\mathrm{cut}} = 6.0$~\AA\ to $4.45$~\AA . The missing contributions to the bond energy energy are compensated by adding an additional attractive term. Here, we use an isotropic embedding term
\begin{equation}
E_{\mathrm{emb}, s} = - \sum_I \sqrt{ \sum_{J\neq I} (a_{\mathrm{emb}})^2 \exp(- b_{\mathrm{emb}} R_{IJ}^2) }
\label{eq:embedding_term}
\end{equation}
motivated by the attractive part of embedded-atom models. The parameters of the embedding term $a_\mathrm{emb}$ and $b_\mathrm{emb}$ are optimized to the hcp, fcc and bcc crystal structures while all other parameters are kept fixed. 

The comparison of Fig.~\ref{fig:H_Urep_Uemb_ini} with Fig.~\ref{fig:H_Urep_ini} shows clearly that the contribution of $s$-electrons in Re-Re interactions can be replaced by a simple embedding term without sizeable loss of model quality. This point of the parameterization protocol concludes with the initial BOP model with a first parameterization of the complete functional form.

\subsubsection{Step 4: From initial BOP to basic BOP}

In this last step, the parameters of all terms, i.e., $H^{\mathrm{ini}}_{IJ}$, $E^{\mathrm{ini}}_{\mathrm{rep}}$ and $E^{\mathrm{ini}}_{\mathrm{emb}}$ are optimized simultaneously to the hcp, fcc and bcc training data. The resulting orthogonal $d$-valent analytic BOP is referred to as basic BOP. The optimization of all BOP parameters leads to very good agreement with the DFT data as shown in Fig. \ref{fig:H_Urep_Uemb}. 
\begin{figure}[ht]
\begin{subfigure}[b]{0.48\columnwidth}
\includegraphics[width=\textwidth]{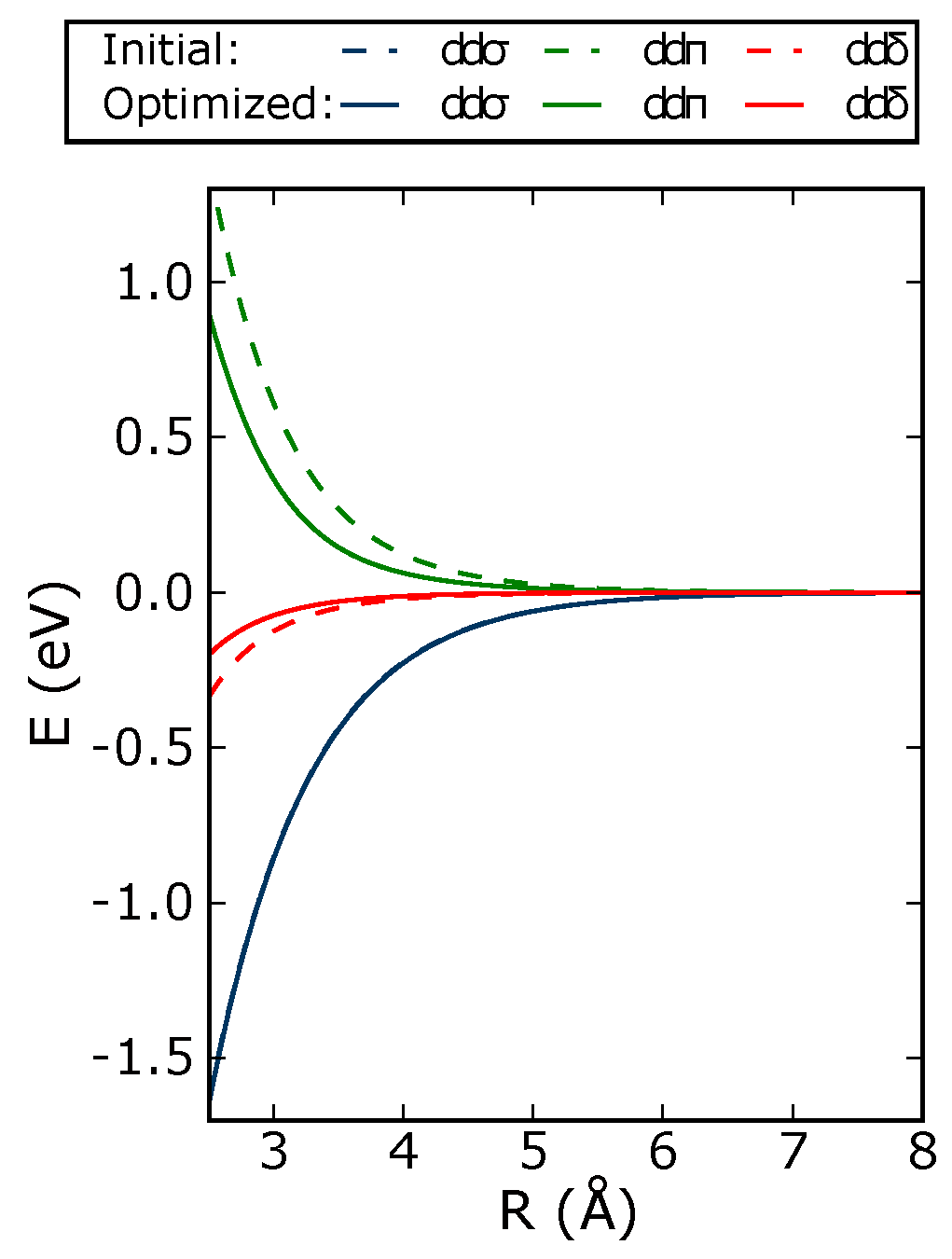}
\caption{a) Matrix elements}
\label{fig:bondints_opt}
\end{subfigure}
\begin{subfigure}[b]{0.48\columnwidth}
\includegraphics[width=\textwidth]{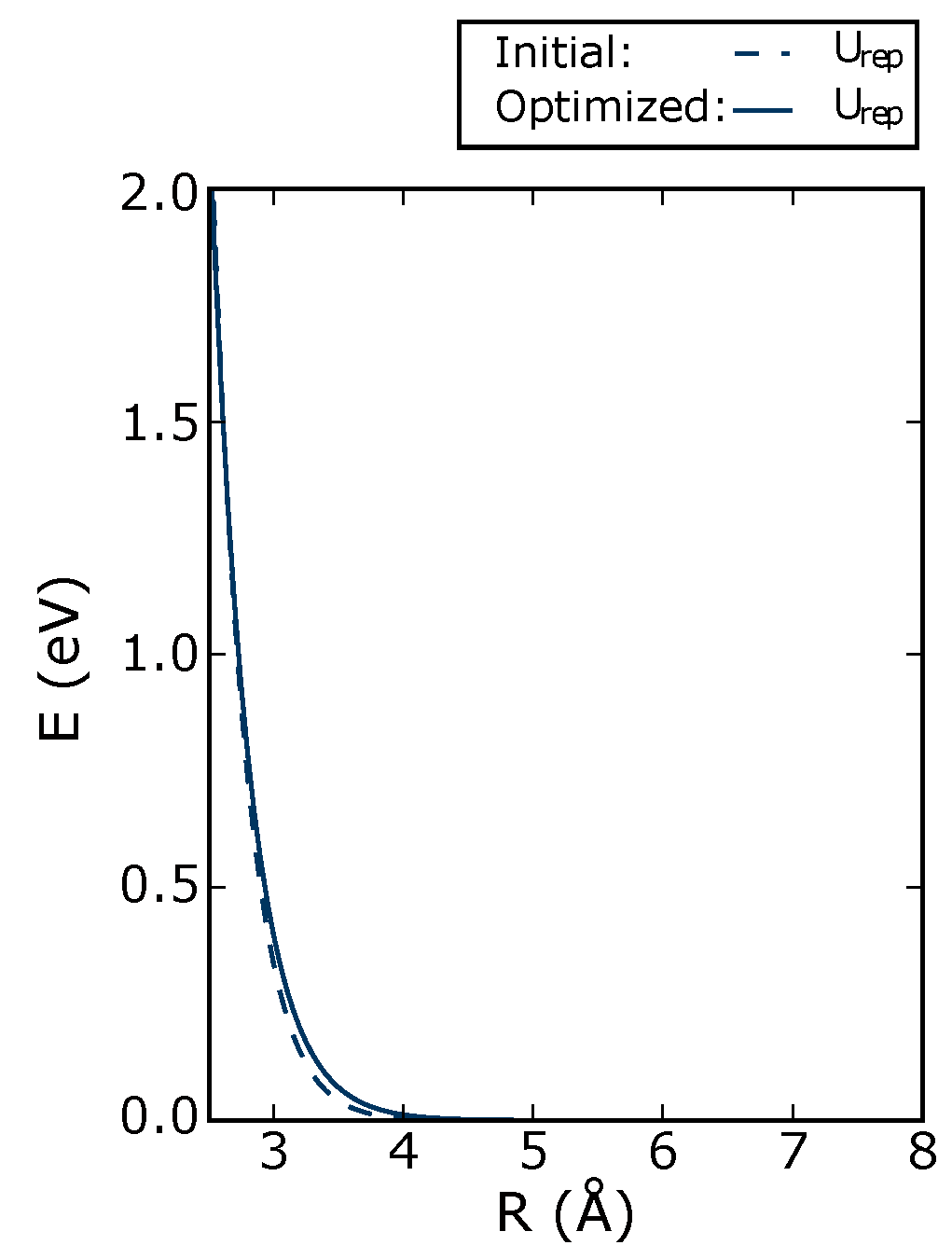}
\caption{b) Pairwise repulsion}
\label{fig:H_Urep_opt}
\end{subfigure}
\caption{Comparison of (a) matrix elements $H_{IJ}$ and (b) pair repulsion $E_{\mathrm{rep}}$ of initial BOP (dashed) and basic BOP (full) obtained by the parameterization protocol for Re hcp, fcc and bcc reference data.}
\label{fig:optimized_model}
\end{figure}
The quality of the initial BOP model becomes apparent by realising that the parameters change only slightly in this last optimization step as shown for the matrix elements and the repulsive energy in Fig.~\ref{fig:optimized_model}. The effective decrease of the range of the matrix elements may be attributed to screening effects that play a role in the bulk reference data but were absent in the downfolding for dimers.

\subsection{\label{sec:transferability_analysis}Transferability analysis of basic BOP}

The basic BOP obtained by the parameterization protocol is optimized for a minimum set of reference data and calls for an assessment of the transferability to other crystal structures and other properties. 
Here, we use 300 random structures that cover the full space of local atomic environments of one atom unit cells in a homogeneous sampling. These structures have been identified earlier in the construction of a map of local atomic environments that is spanned by descriptors based on BOP moments~\cite{Jenke-18}. 

\begin{figure}[ht]
\begin{subfigure}[b]{0.48\columnwidth}
\includegraphics[width=\textwidth]{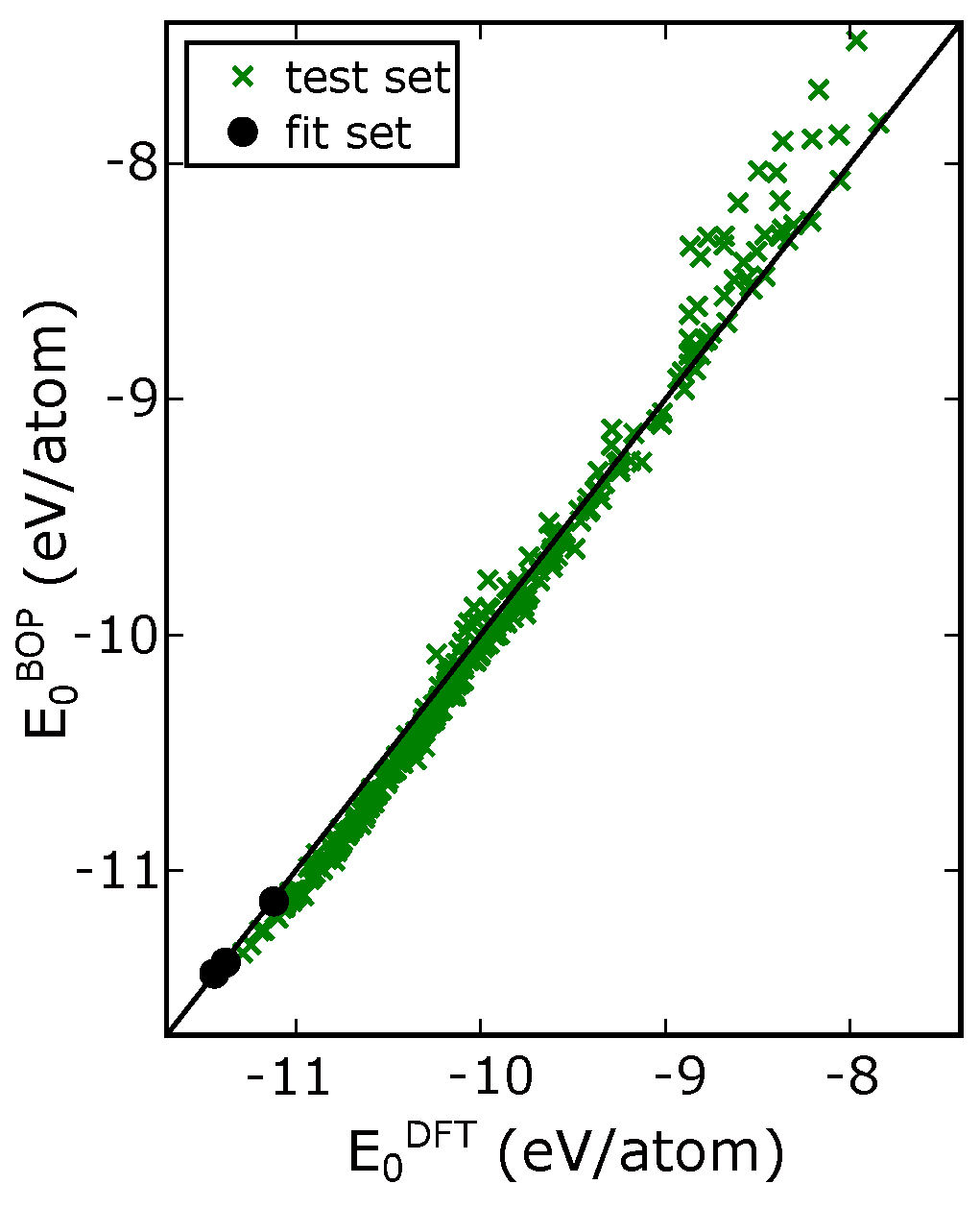}
\caption{a) Equilibrium energy}
\label{fig:initial_model_performance_E0}
\end{subfigure}
\begin{subfigure}[b]{0.48\columnwidth}
\includegraphics[width=\textwidth]{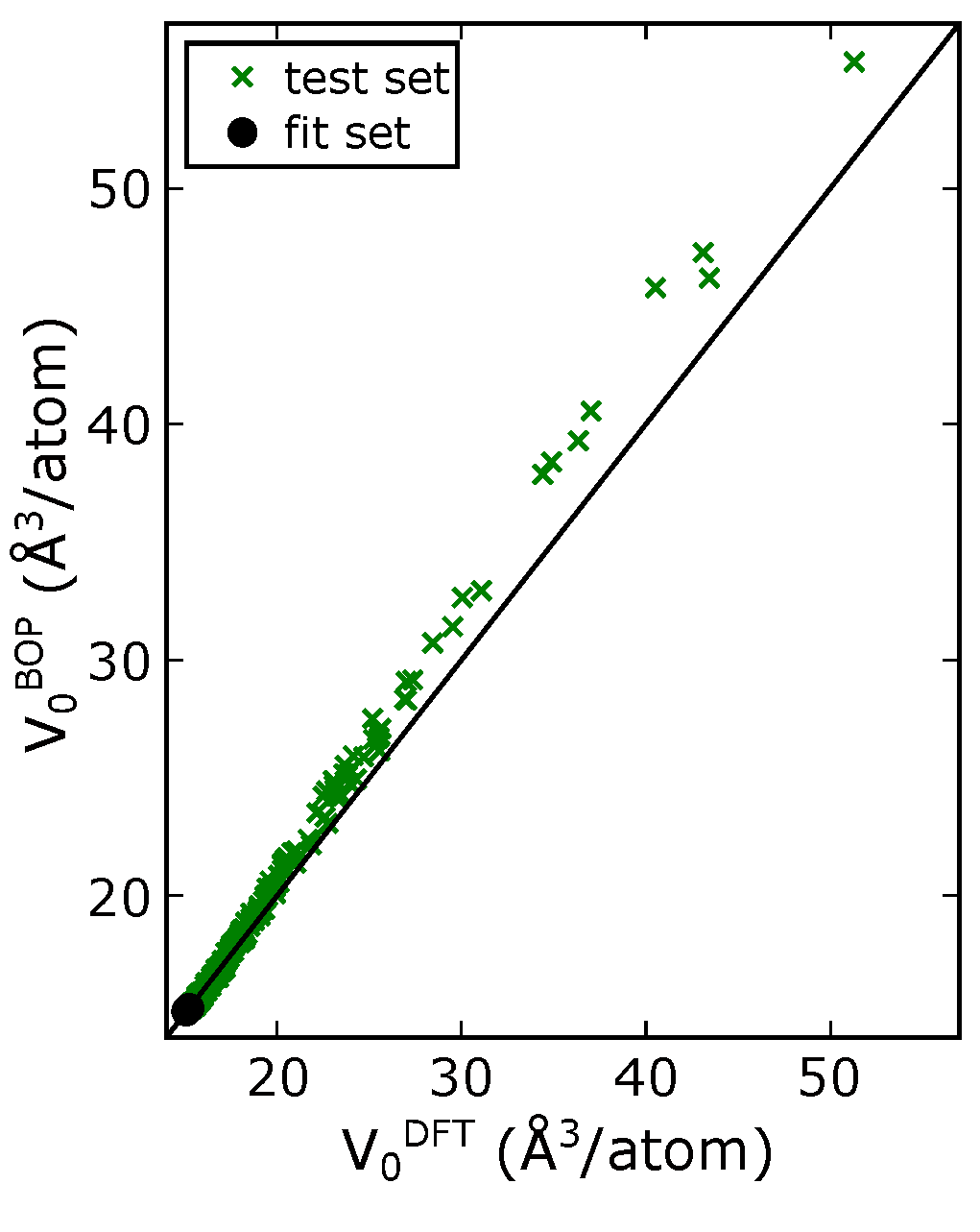}
\caption{b) Equilibrium volume}
\label{fig:initial_model_performance_V0}
\end{subfigure}
\begin{subfigure}[b]{0.48\columnwidth}
\includegraphics[width=\textwidth]{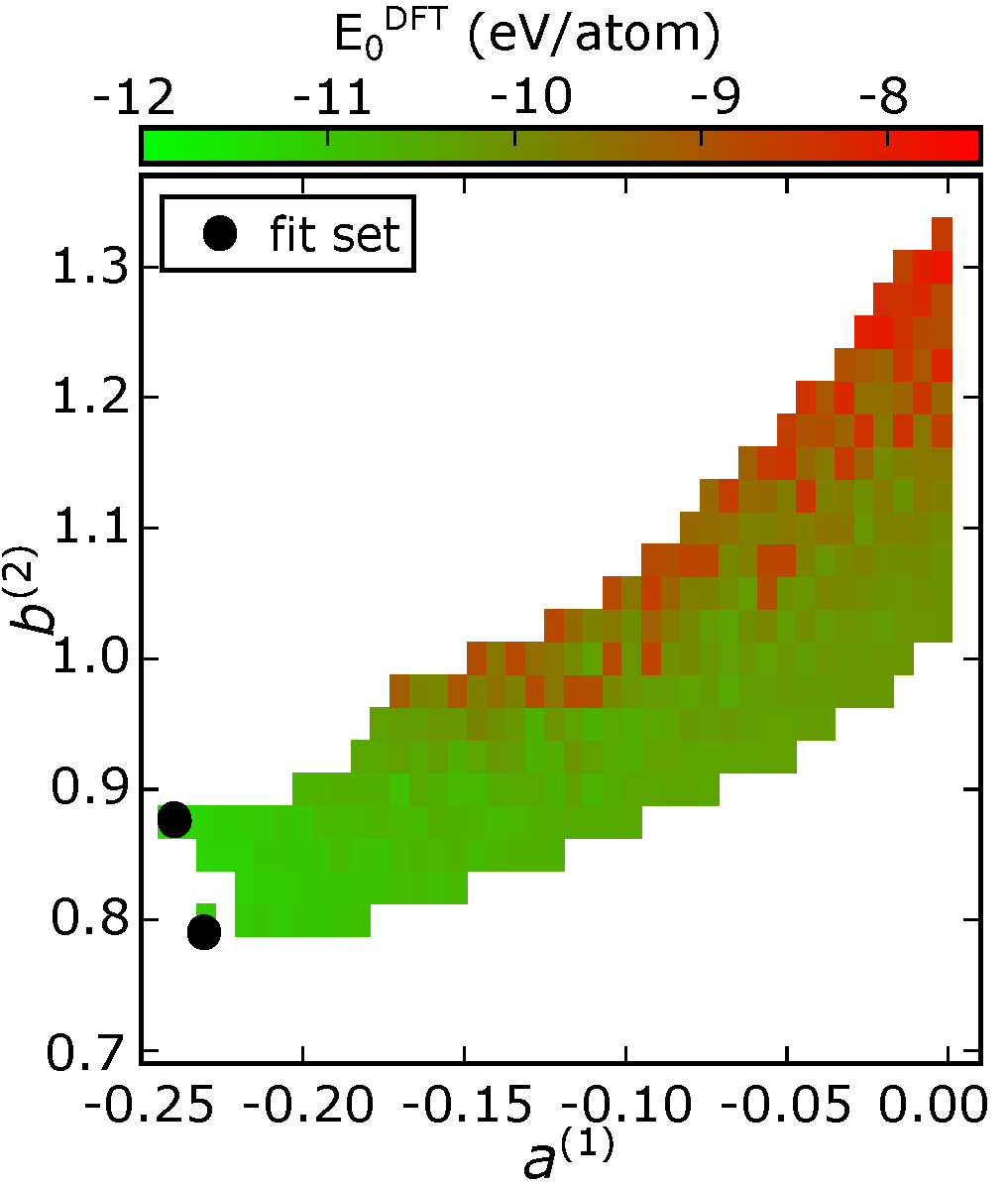}
\caption{c) DFT}
\label{fig:e0_DFT_map_natoms_3}
\end{subfigure}
\begin{subfigure}[b]{0.48\columnwidth}
\includegraphics[width=\textwidth]{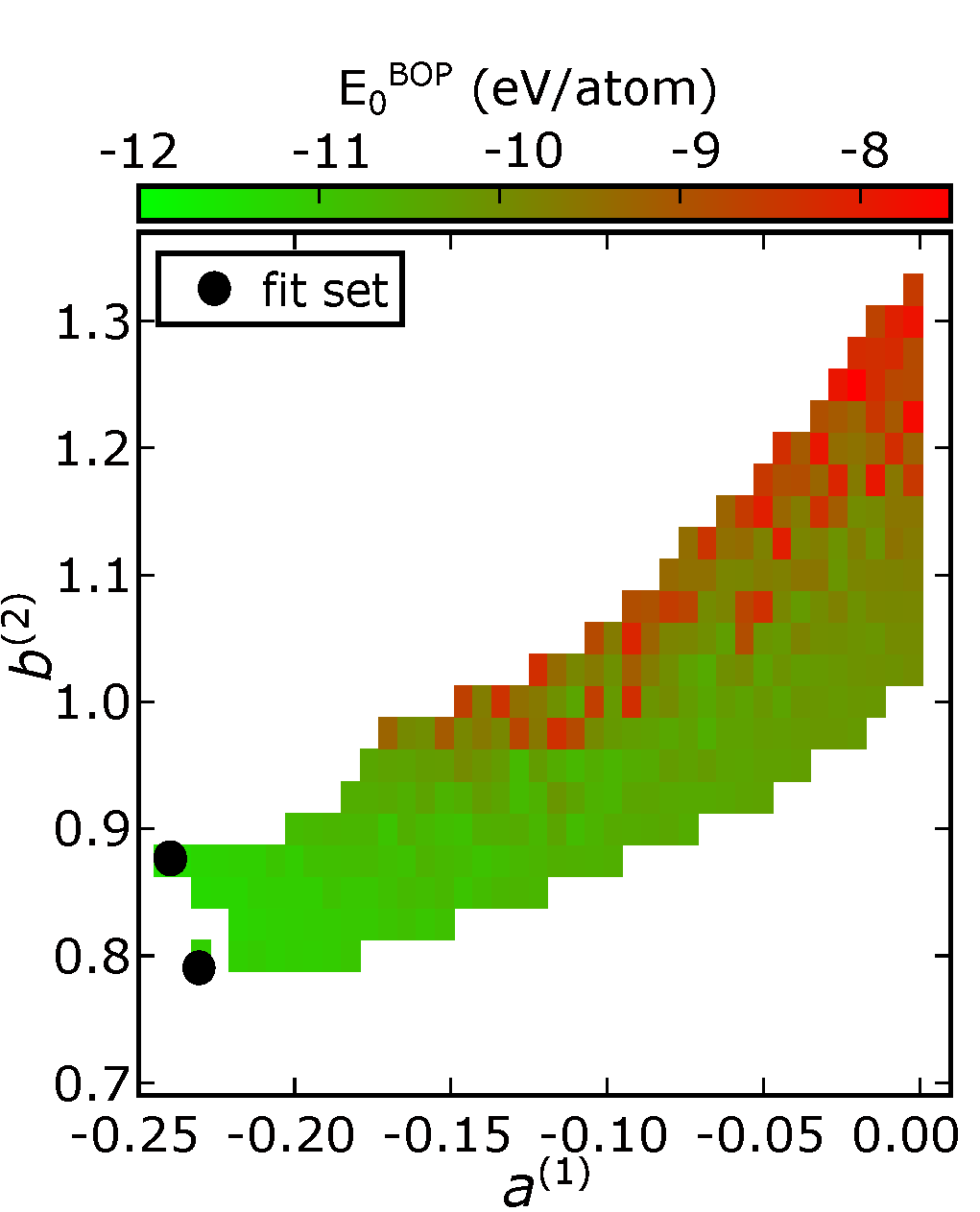}
\caption{d) BOP}
\label{fig:e0_map_natoms_3}
\end{subfigure}
\caption{Transferability of basic BOP to (a) equilibrium energy and (b) equilibrium volume of 300 random structures with 1-atom unit cells as compared to DFT reference data. Variation of equilibrium energy across the complete range of local atomic environments in 1-atom unit cells computed with (c) DFT and (d) basic BOP. The black dots indicate the training data of the basic BOP, i.e. hcp, fcc, and bcc crystal structures.}
\label{fig:initial_model_performance}
\end{figure}

The comparison of the equilibrium energy and equilibrium volume of the random structures predicted by the basic BOP model and the DFT reference data is compiled in Figs.~\ref{fig:initial_model_performance_E0} and \ref{fig:initial_model_performance_V0}. The agreement across the entire range of structures can be considered excellent given that only the energy-volume curves of hcp, fcc, and bcc were used in the parameterization.

The assessments in Figs.~\ref{fig:initial_model_performance_E0} and \ref{fig:initial_model_performance_V0} provide a relation between the basic BOP and the DFT reference data across the range of energies or volumes that is, however, agnostic of the corresponding atomic environments. Therefore we additionally present the equilibrium energies obtained by DFT and basic BOP as color code in the structure similarity map in Figs.~\ref{fig:e0_DFT_map_natoms_3} and \ref{fig:e0_map_natoms_3}, respectively. The coordinates in the map correspond to descriptors based on the moments of the DOS (Eq.~\ref{eq:moments}) from BOP that discriminate different crystal structures~\cite{Turchi-91,cryst6020018} and local atomic environments~\cite{Jenke-18}. The direct relation between the distance of two points in the map that correspond to different crystal structures and the difference in the formation energy of these crystal structures has also been used successfully in machine-learning applications~\cite{Sutton-19}. This analysis shows clearly that the basic BOP captures the equilibrium energy very well across the complete space of local atomic environments of 1-atom unit cells. It is transferable in the region of close-packed crystal structures where it was parameterized but also in regions of open structures with high energies (e.g. simple cubic, 2D square lattice and linear chain at $a^{(1)}=0$, see Ref.~\onlinecite{Jenke-18} for more details). This analysis underlines the intrinsic transferability of even simple BOP models from minimal sets of training data. In the next section we will discuss different strategies to refine this basic BOP to a final BOP for Re.

\section{\label{sec:refinement}Refinement of basic BOP}

\subsection{\label{sec:randomfit} Strategy 1: Towards global transferability}

\begin{figure}[ht]
\begin{subfigure}[b]{0.48\columnwidth}
\includegraphics[width=\textwidth]{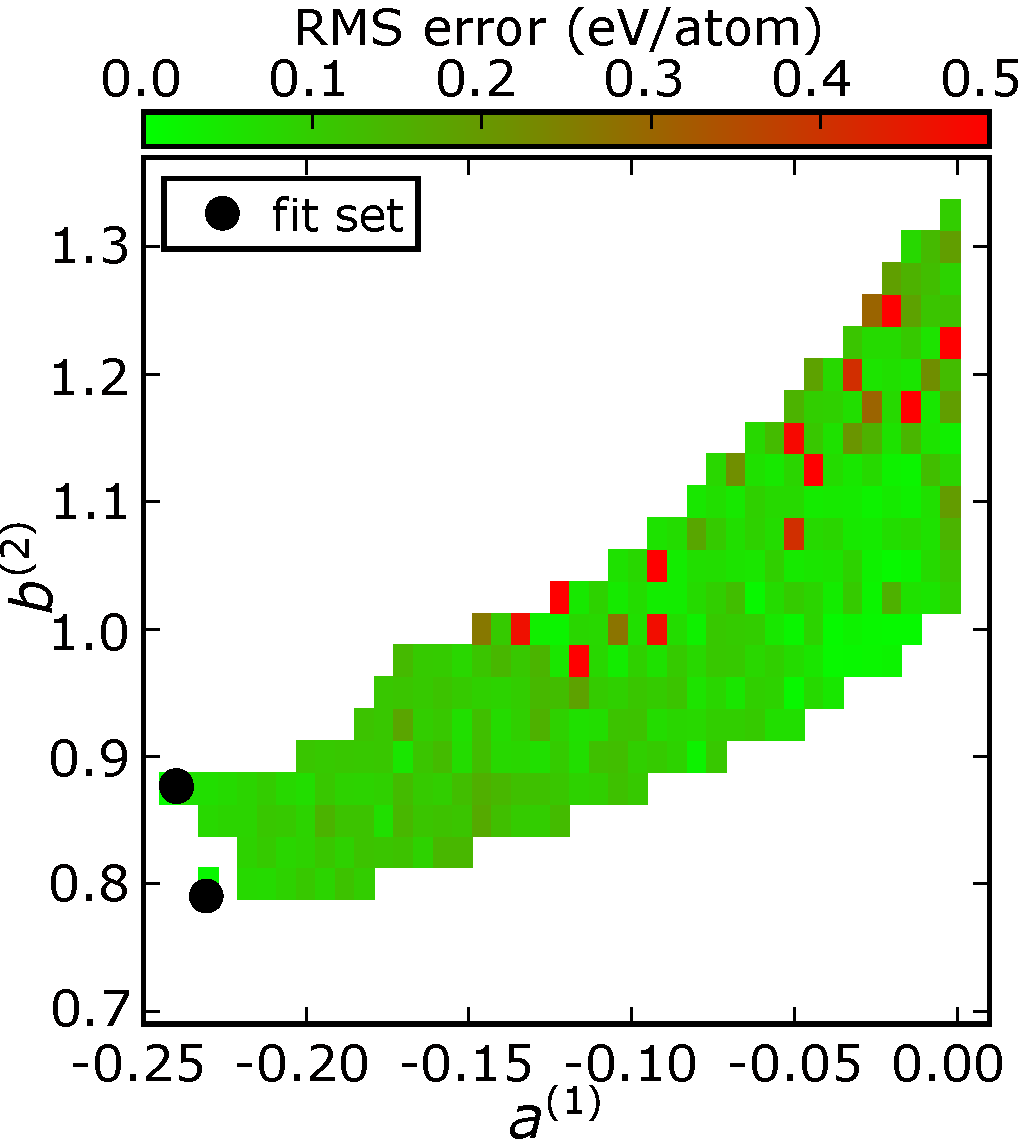}
\caption{a) 3 structures in fit set}
\label{fig:rms_transferability_map_natoms_3}
\end{subfigure}
\begin{subfigure}[b]{0.48\columnwidth}
\includegraphics[width=\textwidth]{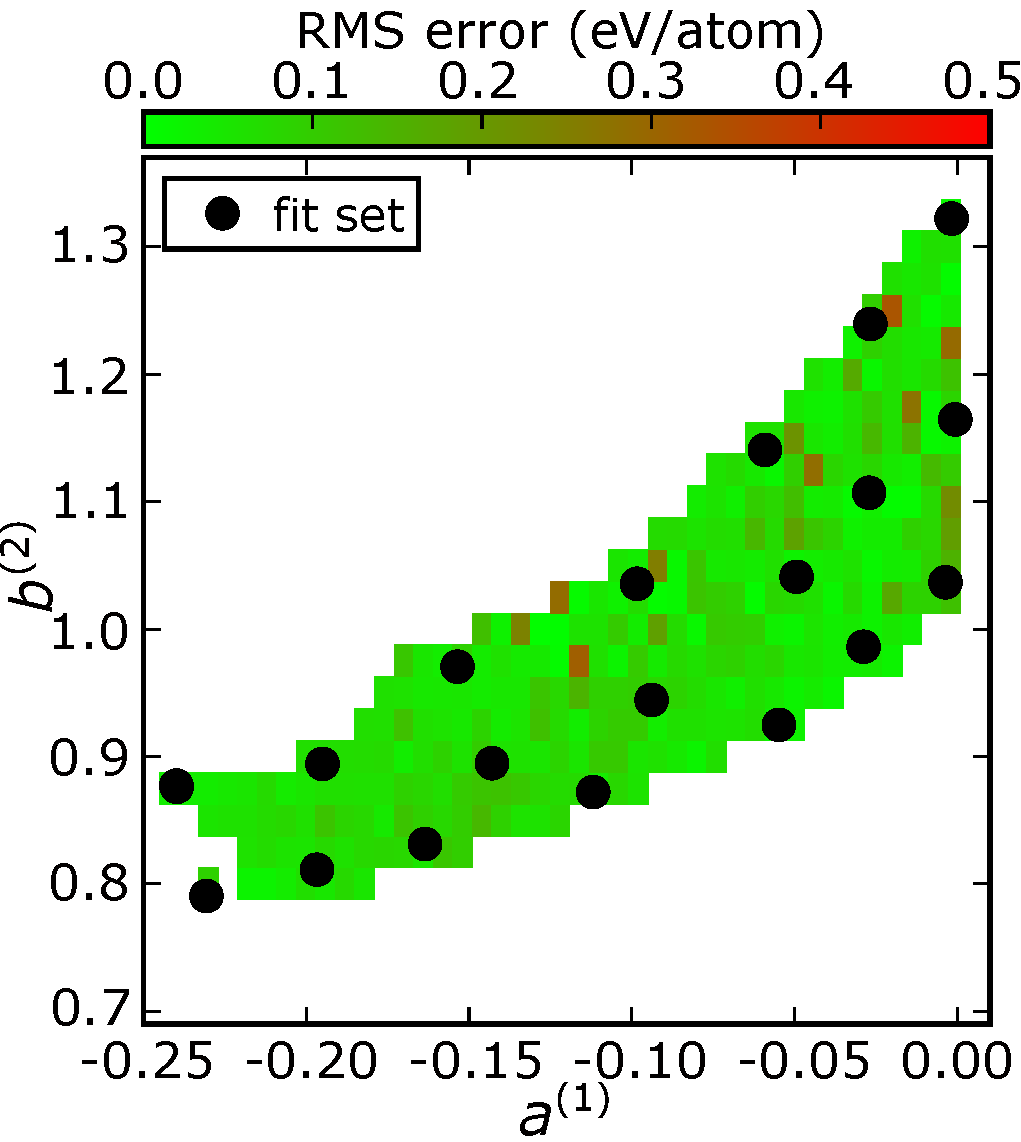}
\caption{b) 20 structures in fit set}
\label{fig:rms_transferability_map_natoms_20}
\end{subfigure}
\begin{subfigure}[b]{0.48\columnwidth}
\includegraphics[width=\textwidth]{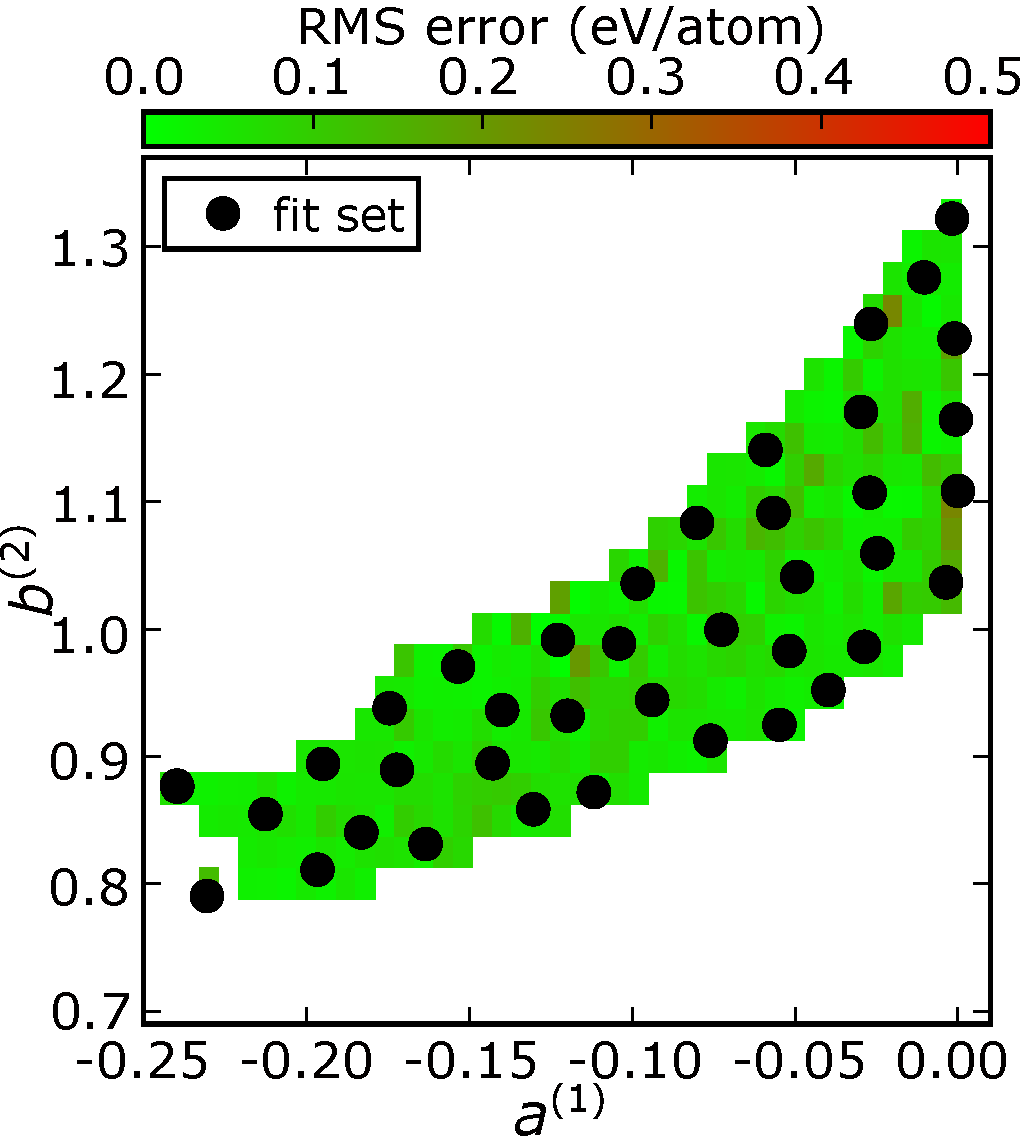}
\caption{c) 40 structures in fit set}
\label{fig:rms_transferability_map_natoms_40}
\end{subfigure}
\begin{subfigure}[b]{0.48\columnwidth}
\includegraphics[width=\textwidth]{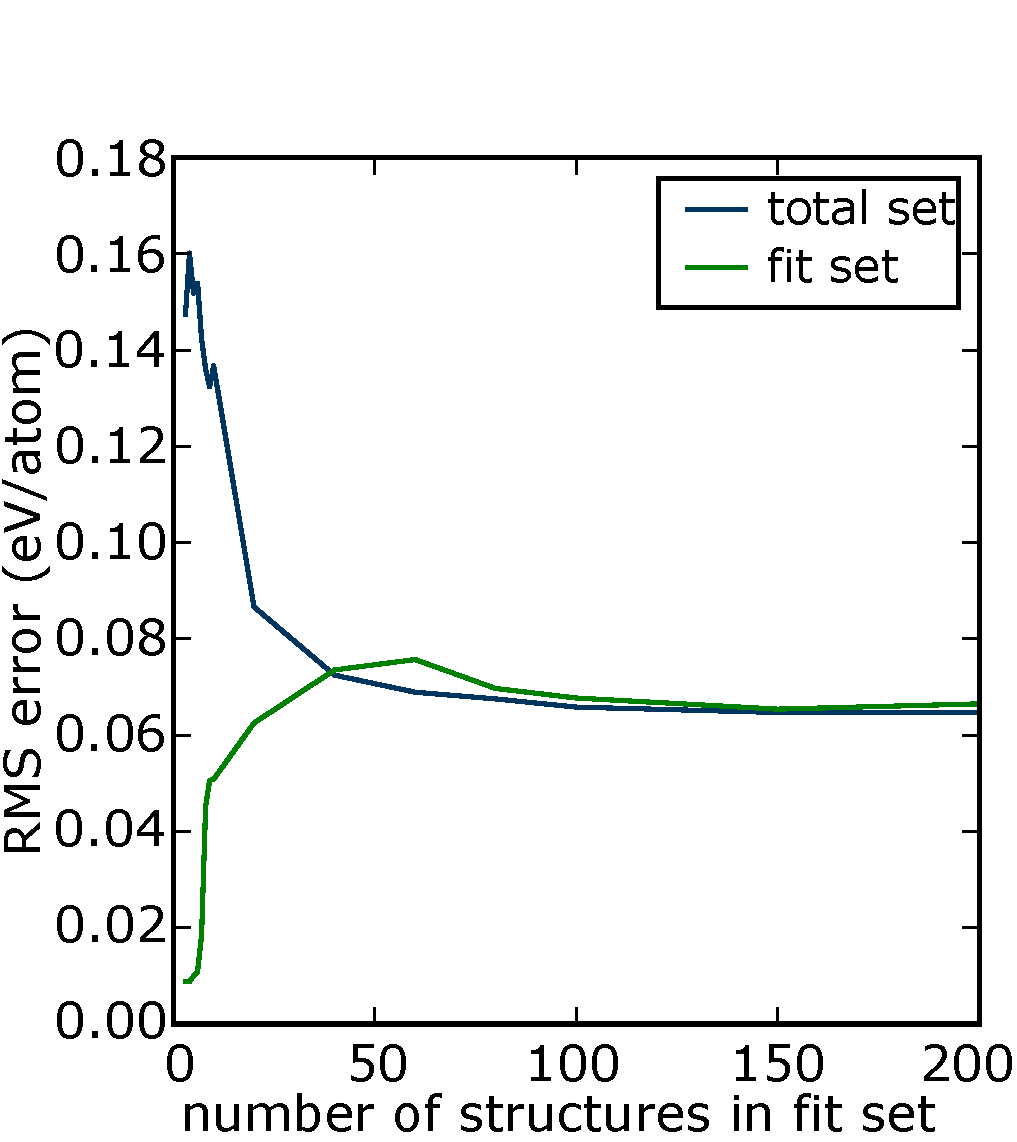}
\caption{d) Learning curve}
\label{fig:learning_curve}
\end{subfigure}
\caption{Refinement of basic BOP model towards global transferability. The RMS error (a) after optimization to hcp, fcc and bcc improves systematically by homogeneous samplings of the space of local atomic environments using (b) 20 and (c) 40 random structures and shows saturation of the (d) learning curve at an RMSE of about 65~meV/atom.}
\label{fig:transferability_improvement}
\end{figure}

In the present parameterization, we deliberately kept the complexity of the BOP model to a minimum. This allows us to perform a transparent analysis of the balance between global transferability and local accuracy that is an inherent compromise in many developments of interatomic potentials. With strategy 1, we demonstrate a route to a refinement of the basic BOP towards global transferability. To this end we use the map of local atomic environments and analyse the RMS error in the equilibrium energy that we obtain with having used only hcp, fcc and bcc structures in the training data. The distribution of the RMSE shows that the basic BOP has less transferability for structures with large distance to the fit set, see Fig.~\ref{fig:rms_transferability_map_natoms_3}. In other words, the map quantifies and confirms graphically the expectation that larger errors in energy that are to be expected for further extrapolations from the training data used in the parameterization.

An apparent strategy to improve the global transferability is to extend the set of training data. The map of local atomic environments offers an access to systematically carry out homogeneous and extensive samplings of the full range of local atomic environments.
The basic BOP is refined by optimizing all parameters to extended sets of training data. In particular, we add the energy-volume curves of random structures that are selected to achieve homogeneous samplings of the range atomic environments with increasing density. In Figs.~\ref{fig:rms_transferability_map_natoms_20} and \ref{fig:rms_transferability_map_natoms_40}, we see that the RMS error computed for all 300 structures is systematically reduced by successively extending the training data to 20 and 40 random structures. Repeating the refinements of the BOP for homogeneous samplings with up to 200 random structures leads to the learning curves shown in Fig.~\ref{fig:learning_curve}. We find that 40 random structures are already a good representation of the range of atomic environments of 1-atom unit cells. The unusual behaviour of a crossing of the two curves and a higher RMS error in the fit set than in the total set of structures is an artefact of the special choice of the reference data. The learning curve converges to an RMS error of about 65~meV/atom which corresponds to 1.8\% of the energy range of the considered structures.
These results demonstrate the iterative optimization of transferability to 1-atom unit cells by systematic samplings of local atomic environments. Despite its benefits, this approach requires further work towards sufficiently complete sets of complex unit cells to cover the atomic environments relevant for other crystal structures (e.g. TCP phases), defects (e.g. vacancies) and property calculations (e.g. displacements for elastic constants).

\subsection{Strategy 2: Towards local accuracy}

With the second strategy we demonstrate a refinement of the basic BOP towards the description of specific properties without actively enforcing global transferability as in the first strategy. Here, we choose the target properties as TCP phases and elastic properties motivated by typical applications of Re. The training data consists of hcp, fcc, bcc, A15, C15, and $\sigma$ phases that ensure a certain variety of local atomic environments in terms of the 12-, 14-, 15- and 16-fold coordination polyhedra of the nearest-neighbour shells in these structures. It furthermore contains the energies of elastic deformations of the ground-state hcp structure. The remaining reference data (cf. Sec.~\ref{sec:referencedata}) is used for testing the model. 

An integral part of this optimization strategy is an appropriate balancing of the weights of target properties in the cost function (Eq.~\ref{eq:cost_function}). 
For the energy-volume data of different crystal structures, we allow higher errors for structures that are energetically less favorable than the hcp ground-state using 
\begin{equation}\label{eq:weighting}
w_{\mathrm{struc}} = \exp \left(\frac{E_{\mathrm{hcp}}^{(0)} - E_{\mathrm{struc}}^{(0)}}{0.1 \mathrm{eV}} \right)
\end{equation}
with the respective equilibrium energies per atom $E^{(0)}_{\mathrm{struc}}$. The denominator of $0.1 \ \mathrm{eV}$ is chosen empirically and corresponds to a temperature of about 1200~K. 
For the elastic properties, we introduce weights of 
\begin{equation}
w_{\mathrm{ela}} = a_{\mathrm{ela}} \frac{\Delta E^{\mathrm{struc}}}{\Delta E^{\mathrm{ela}}},
\label{eq:weight_ela}
\end{equation}
to adjust the order of magnitude of the largest energy difference of the energy-volume curves
$\Delta E^{\mathrm{struc}}$ ($\approx$ 1~eV/atom) and the largest energy difference of the elastic deformations $\Delta E^{\mathrm{ela}}$ ($\approx$ 50~meV/atom). 
Without this adjustment, the relative accuracy of the energy-volume curves will be higher than the elastic deformations due to their different range of energy.

For our assessment of the balance between global transferability and local accuracy, we use different values of $a_{\mathrm{ela}}$ in Eq.~\ref{eq:weight_ela} that range from 0 (elastic deformations disregarded) to 100 (elastic deformations dominate optimization). For each value, we perform a set of optimizations to the 190 data points of the training data that start from 30 different initial guesses. The latter are generated by randomly changing the 23 parameters of the basic BOP model within a Gaussian distribution with a width of 5\% of the initial parameter value. With the sets of randomized initial guesses we enable the downhill optimization algorithm to detect more than only one local minimum in the high-dimensional parameter space.

\begin{figure}[ht]
\includegraphics[width=\columnwidth]{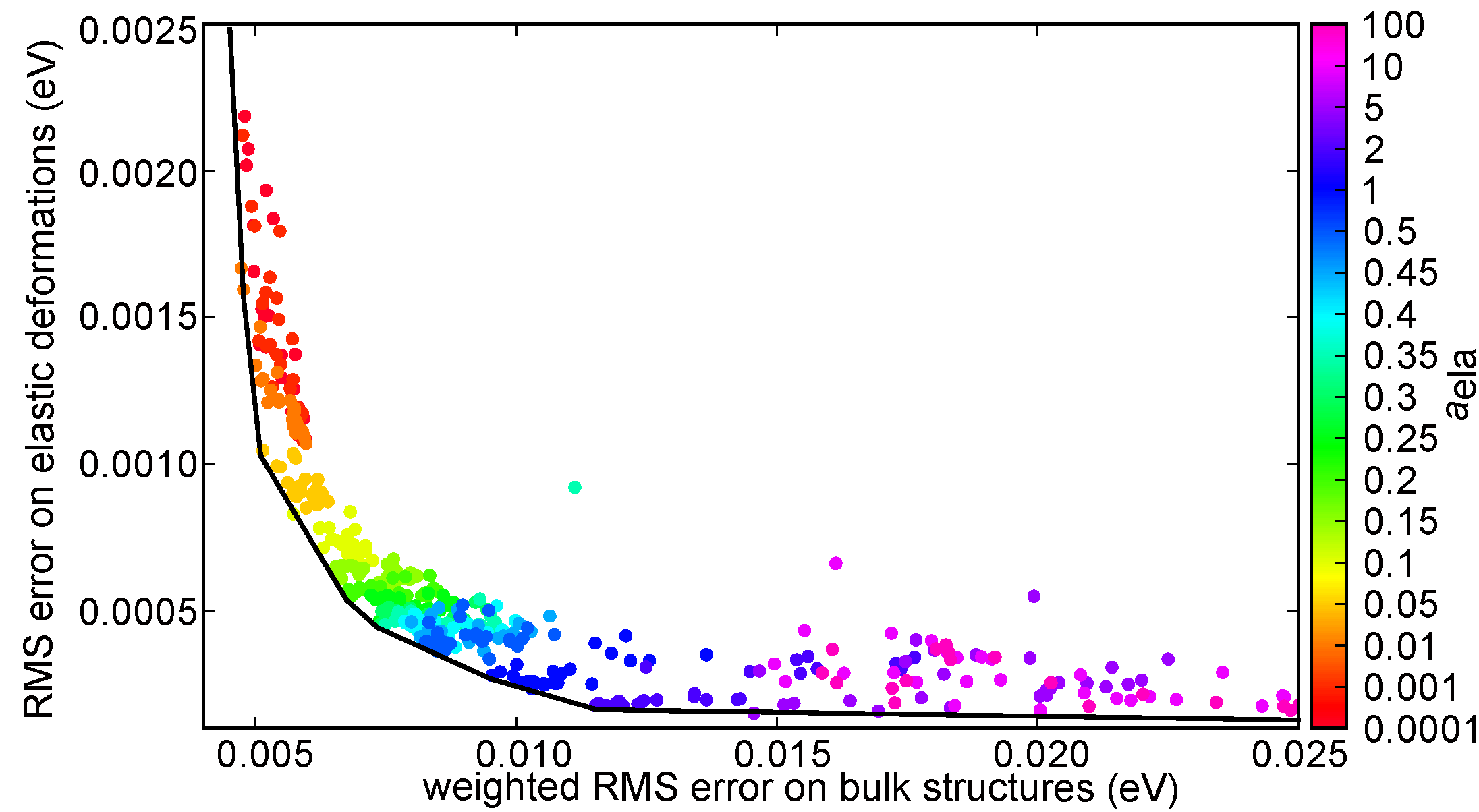}
\caption{RMS errors of bulk structures and of elastic properties of different BOP models (points) optimized with different weights $a_{\mathrm{ela}}$ of Eq.~\ref{eq:weight_ela} (different colors), each starting from a set of different initial guesses (points of same color). The Pareto front (black line) indicates the balance of capturing these different properties.}
\label{fig:Pareto_front}
\end{figure}

Each of these optimizations leads to a BOP model that we assess with respect to the RMS errors of bulk structures and elastic properties comprising the training data as shown in Fig.~\ref{fig:Pareto_front}. Overall, we observe the expected existence of a Pareto front. This marks the limit of optimizing one set of properties without compromising the other set, here bulk structures versus elastic deformations or vice-versa. The close proximity of nearly all BOP models to the Pareto front demonstrates the overall robustness of the minimization procedure. The remaining scatter at the Pareto front confirms the existence of multiple local minima in the high-dimensional parameter space. 
The excellent capturing of bulk structures or elastic deformations for low or high values of $a_\mathrm{ela}$, respectively, shows the effect of our training data weighting with Eq.~\ref{eq:weight_ela}. The minimum RMSE for bulk structures is less then 5~meV/atom, for elastic deformations less than 0.1~meV/atom. The comparably high quality for bulk structures and elastic deformations across the whole Pareto front is an indicator of the intrinsic transferability of the BOP models. This analysis shows that the refinement of the basic BOP model can be systematically targeted to a specific balance of the local accuracy for specific properties. The selection of one final BOP model from this set of candidate BOP models by additional criteria is discussed in the following.

\section{\label{sec:selection}Selection of final BOP}

\subsection{\label{sec:criteria}Selection criteria}

For the selection of one model from the different optimizations in Fig.~\ref{fig:Pareto_front}, we specify further criteria that are formulated as tests motivated by the material system Re. The numerical values of the quantitative tests are chosen empirically such as to minimize the number of models that pass all tests. We emphasize that different material systems and different applications will need qualitatively and quantitatively different tests. For Re in this work, the following tests are performed:
\begin{itemize}
\item \textbf{Test 1}: error in bulk structures below 0.05~eV or error in energy difference to hcp below 40\% \item \textbf{Test 2}: error in elastic constants below 90~GPa
\item \textbf{Test 3}: hcp and dhcp structures correctly ordered
\item \textbf{Test 4}: error in $c/a$ ratio of hcp below 0.01
\item \textbf{Test 5}: error in vacancy formation-energy below 0.8~eV
\item \textbf{Test 6}: error in vacancy diffusion-barrier within and in between basal planes below 0.5~eV
\item \textbf{Test 7}: ISF formation energy larger than 20 $\mathrm{mJ}/\mathrm{m}^2$
\item \textbf{Test 8}: error in SIA formation energies below 2.2~eV
\end{itemize}

\begin{figure}
\begin{subfigure}[b]{0.48\columnwidth}
\includegraphics[width=\textwidth]{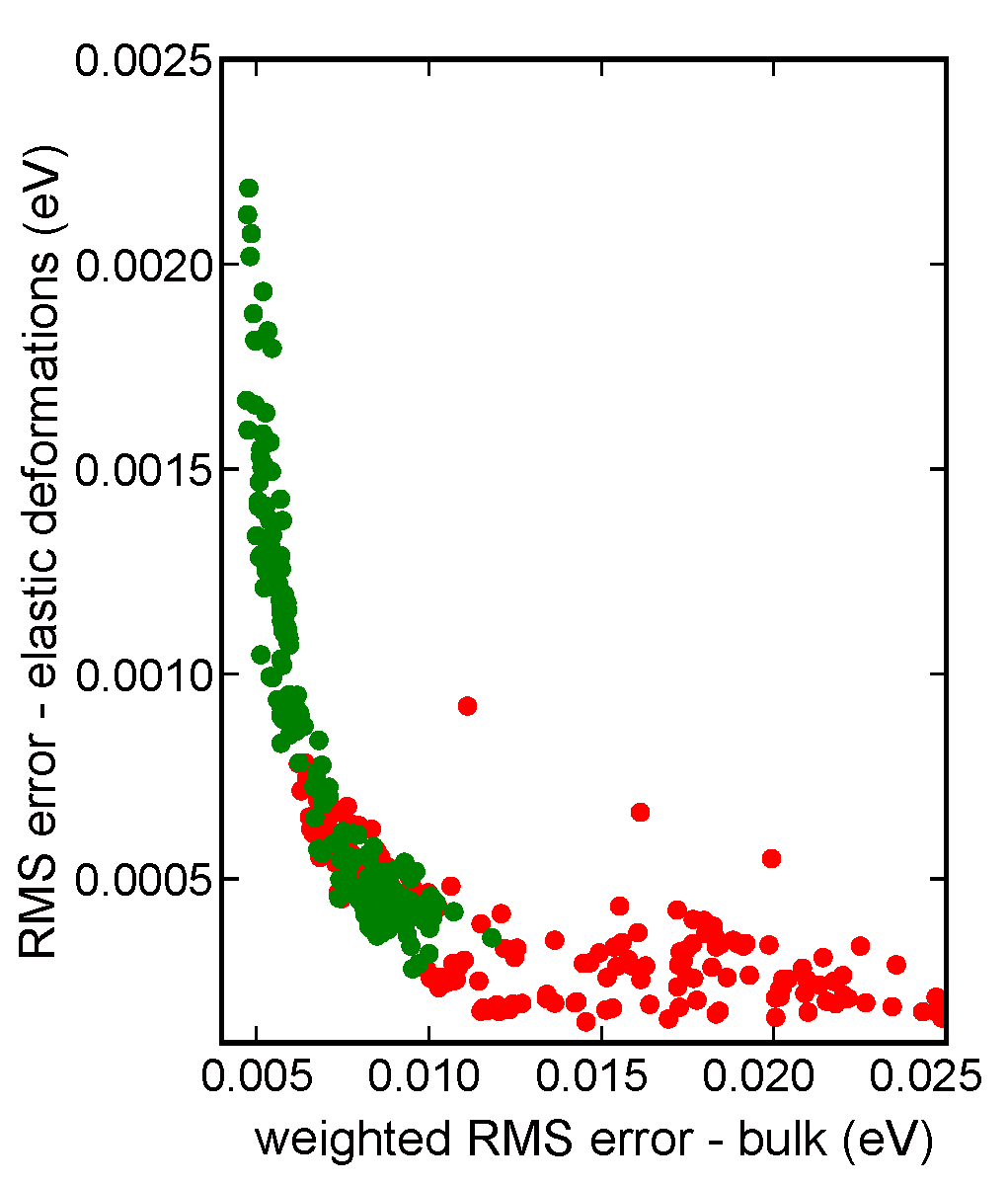}
\caption{Test 1: bulk structures}
\label{fig:test4}
\end{subfigure}
\begin{subfigure}[b]{0.48\columnwidth}
\includegraphics[width=\textwidth]{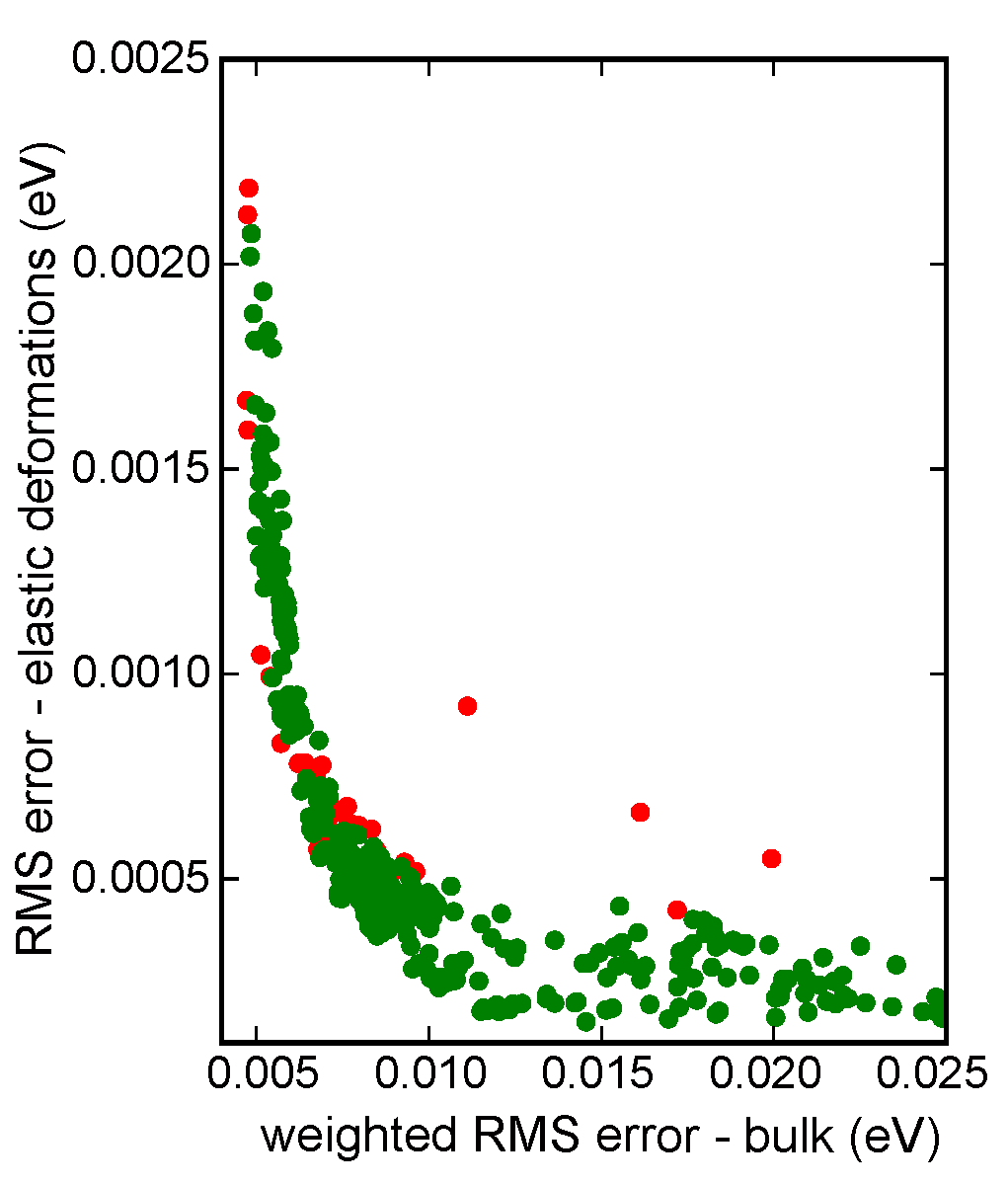}
\caption{Test 2: elastic constants}
\label{fig:test3}
\end{subfigure}
\begin{subfigure}[b]{0.48\columnwidth}
\includegraphics[width=\textwidth]{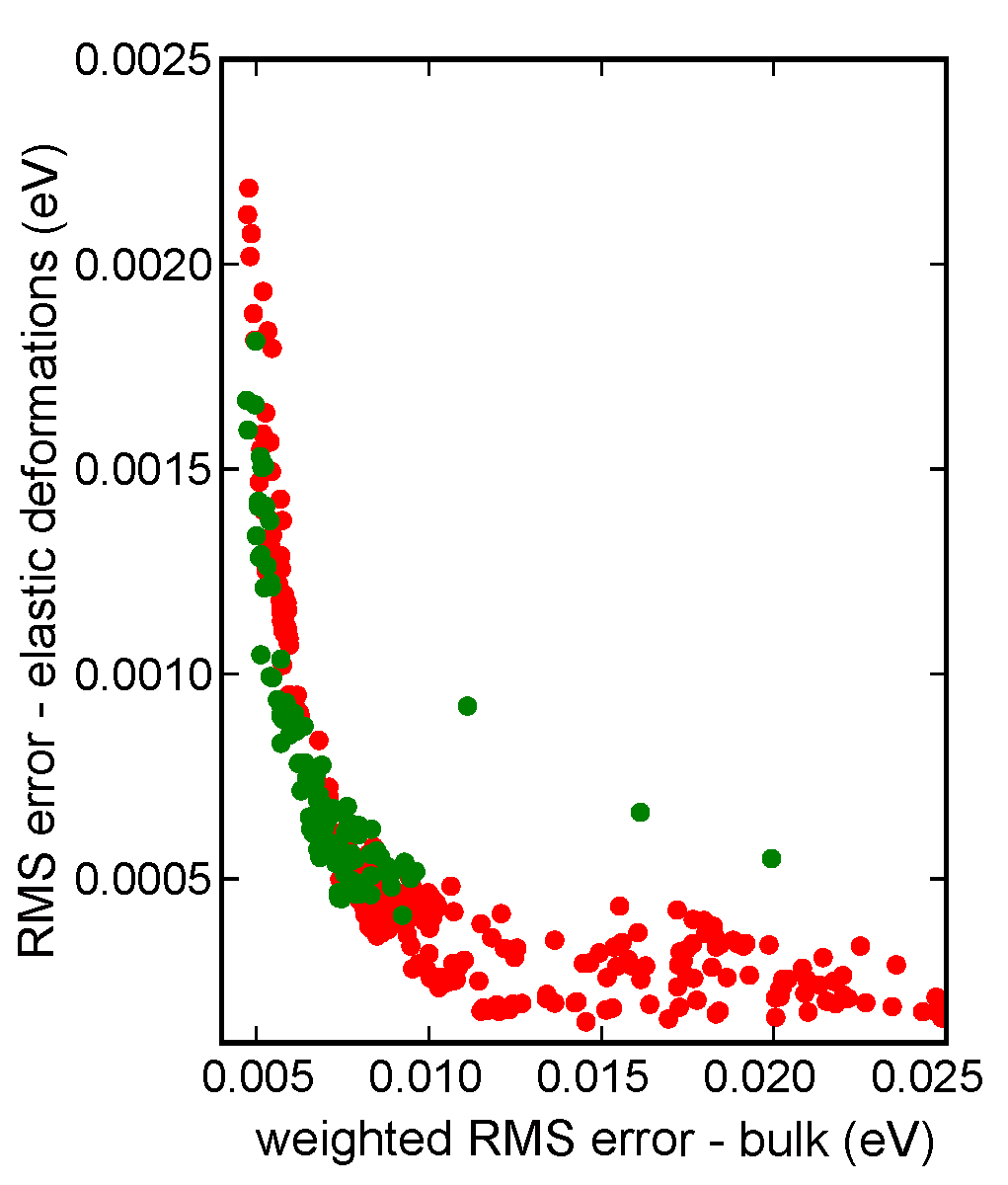}
\caption{Test 3: hcp/dhcp ordering}
\label{fig:test0}
\end{subfigure}
\begin{subfigure}[b]{0.48\columnwidth}
\includegraphics[width=\textwidth]{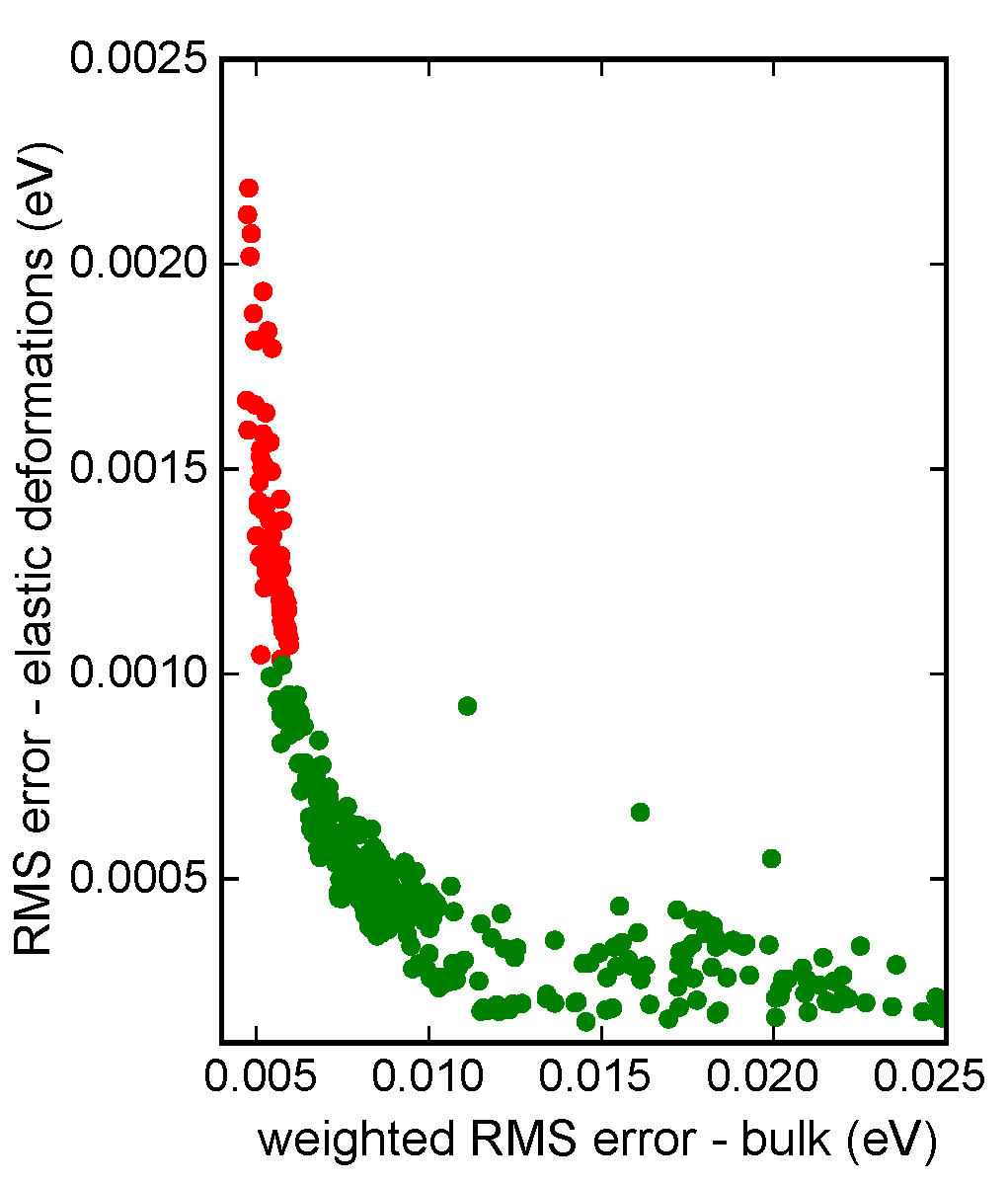}
\caption{Test 4: $c/a$ ratio}
\label{fig:test5}
\end{subfigure}
\begin{subfigure}[b]{0.48\columnwidth}
\includegraphics[width=\textwidth]{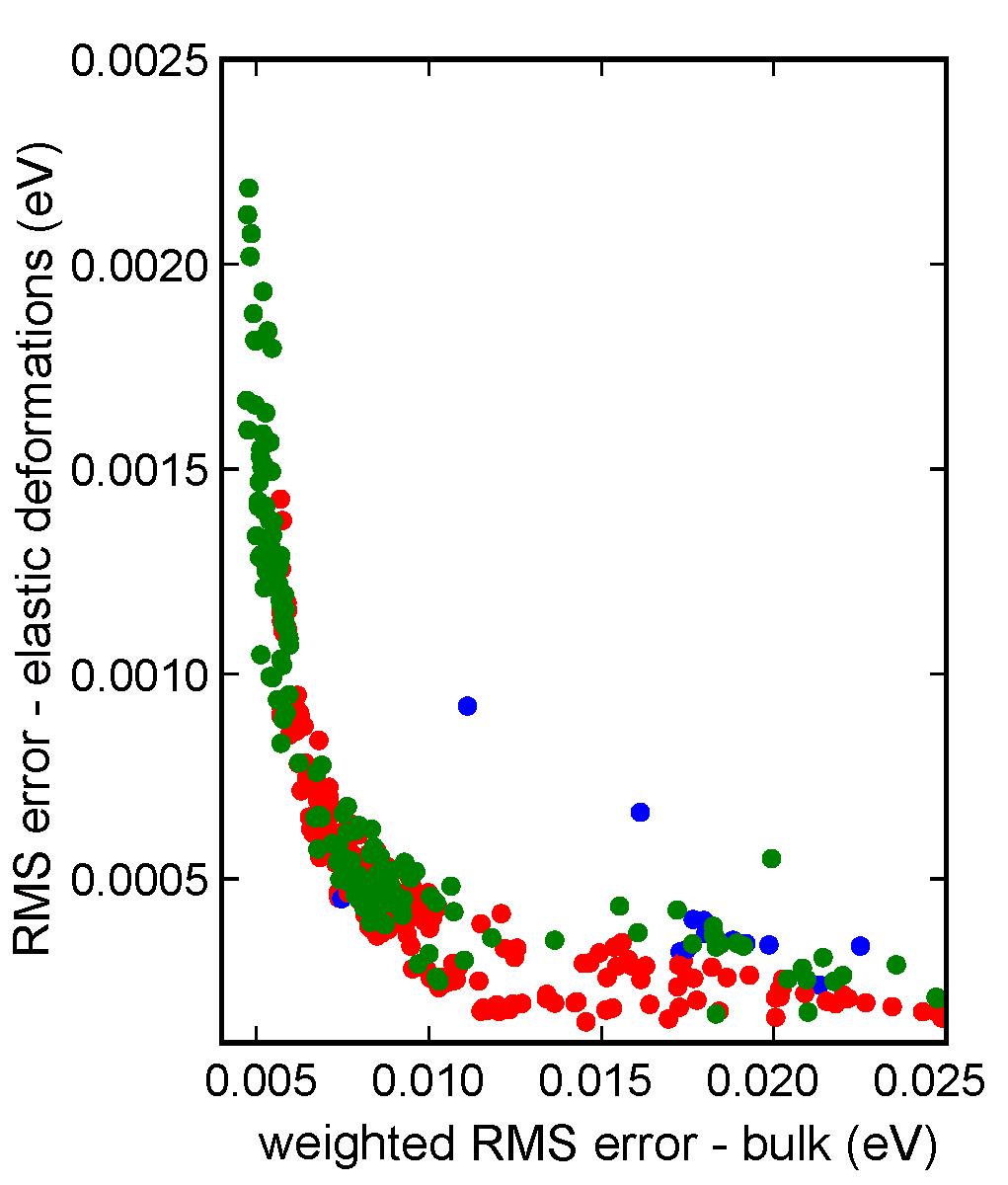}
\caption{Test 5: vacancy formation}
\label{fig:test1}
\end{subfigure}
\begin{subfigure}[b]{0.48\columnwidth}
\includegraphics[width=\textwidth]{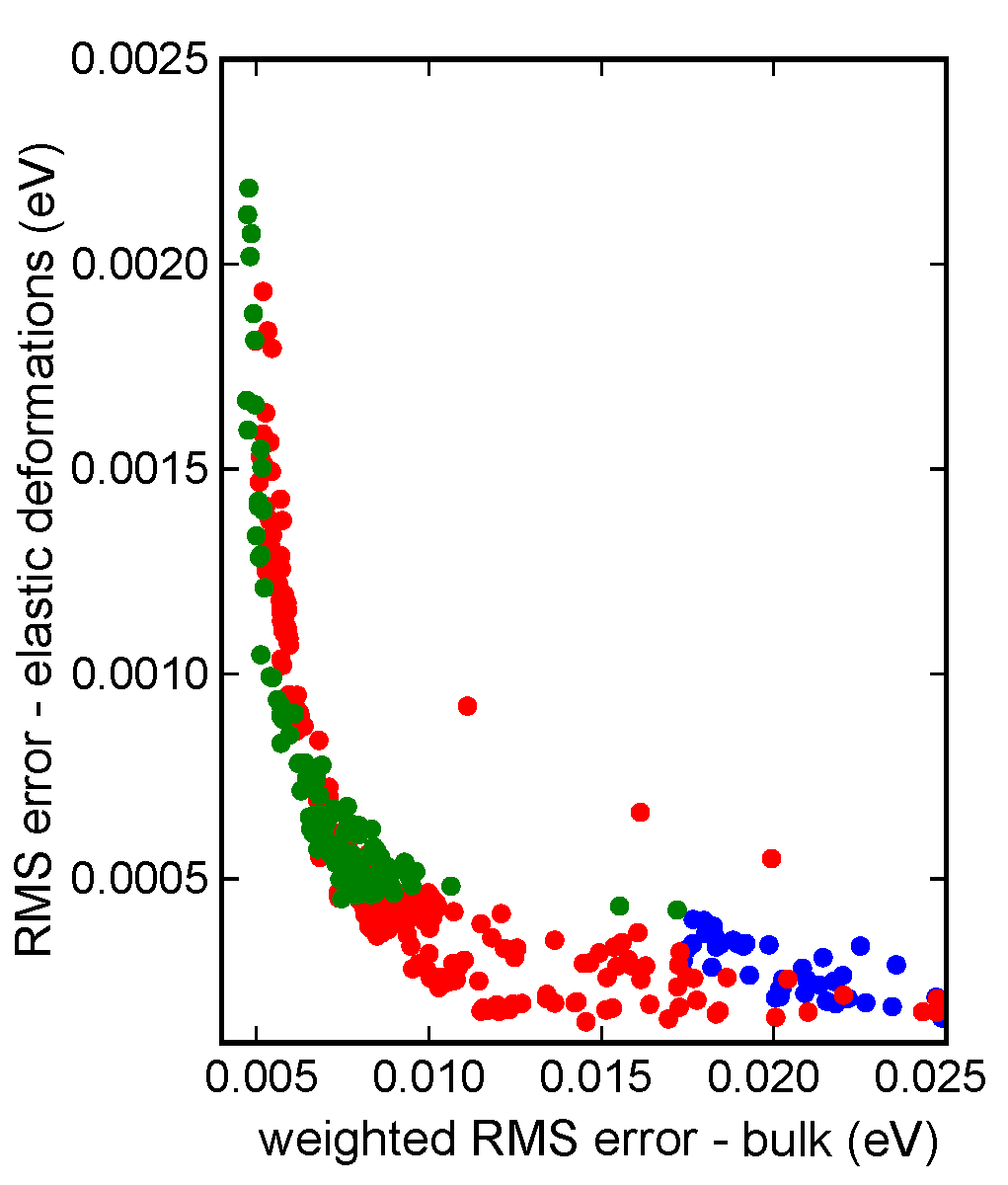}
\caption{Test 6: vacancy diffusion}
\label{fig:test2}
\end{subfigure}
\begin{subfigure}[b]{0.48\columnwidth}
\includegraphics[width=\textwidth]{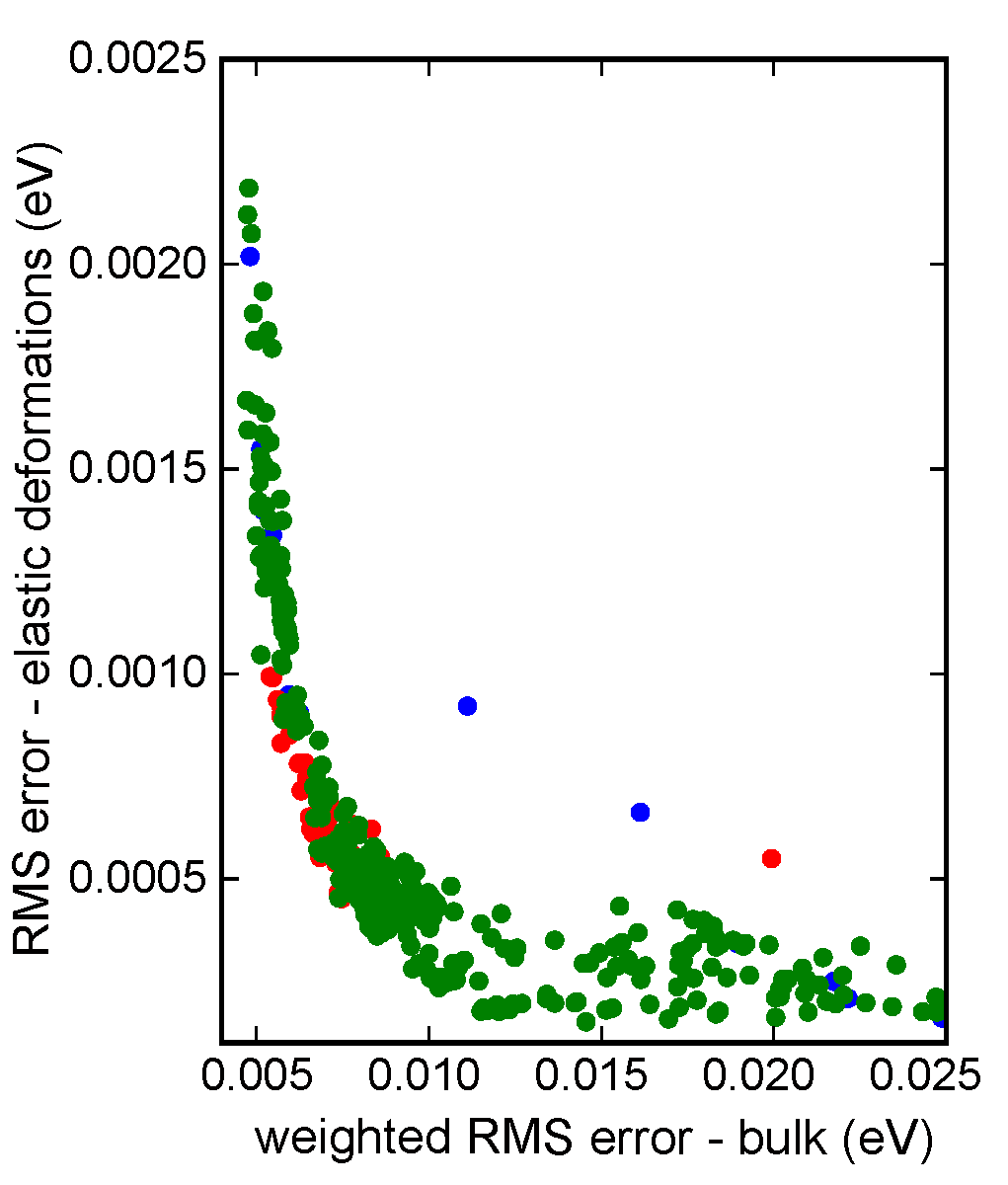}
\caption{Test 7: ISF energy}
\label{fig:test6}
\end{subfigure}
\begin{subfigure}[b]{0.48\columnwidth}
\includegraphics[width=\textwidth]{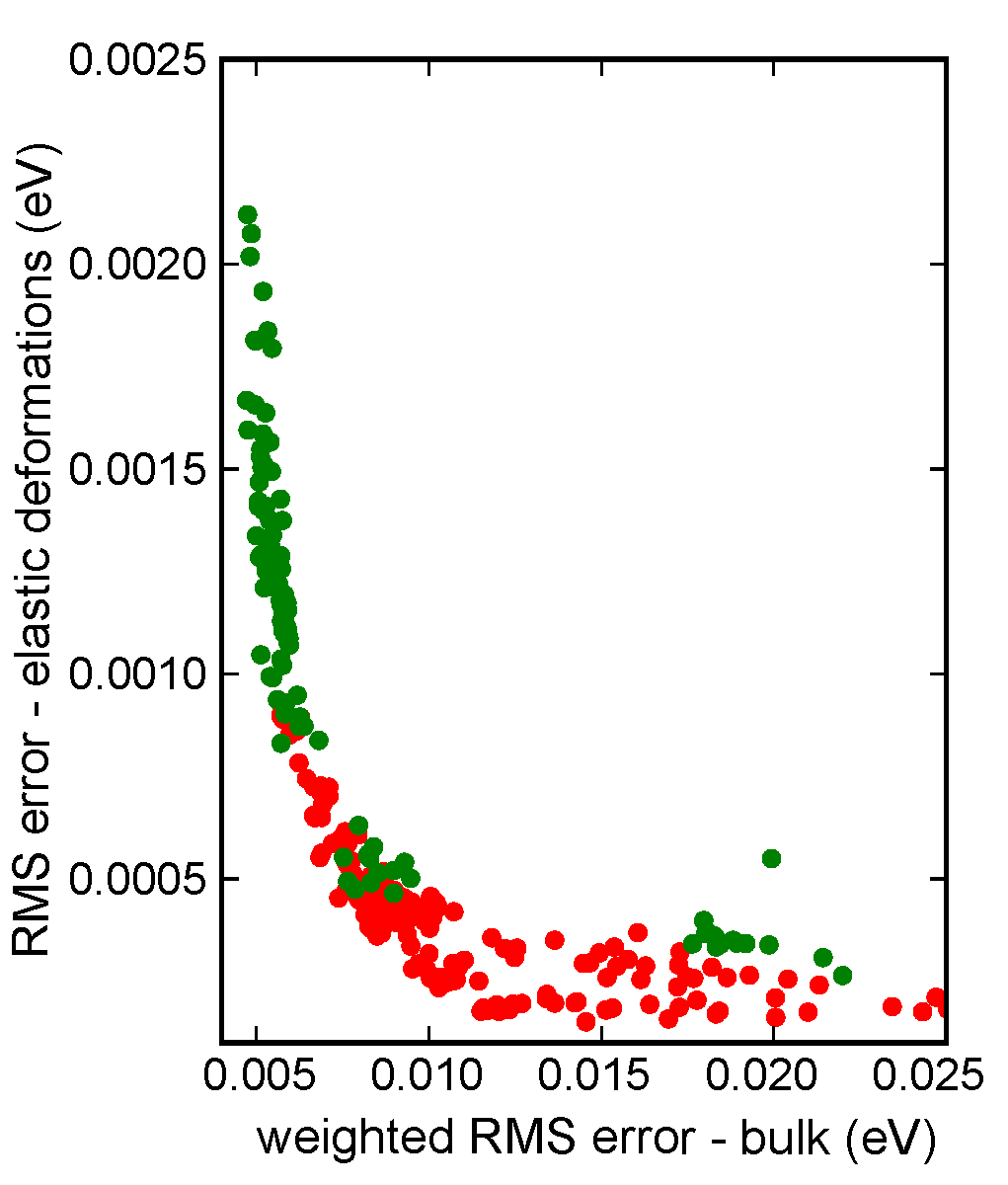}
\caption{Test 8: SIA energy}
\label{fig:test7}
\end{subfigure}
\caption{Performance of BOP models from Fig.~\ref{fig:Pareto_front} on test 1-8 with green/red color signalling passed/failed tests. The points marked in blue correspond to the calculations which exceeded a maximal run time with respect to atomic relaxation.}
\label{fig:test}
\end{figure}

The performance of all BOP models of Fig.~\ref{fig:Pareto_front} on the individual tests is compiled in Fig.~\ref{fig:test}.
Test 1 on the bulk structures is passed for models near the Pareto front with RMS error on bulk structures of less than about 10~meV/atom. 
Test 2 on elastic constants is passed by the majority of models across a broad range of RMSE values.
Test 3 of the dhcp/hcp energy difference is passed by less models than test 2 on bulk structures as it requires to resolve the hcp/dhcp energy difference of only 1.76~meV/atom in DFT.  
Test 4 on the $c/a$ ratio is passed if the elastic deformations around the equilibrium are captured with sufficient accuracy. 
Test 5 on vacancy formation is passed for several models over a broad range of weights. 
Test 6 is passed by less models than test 5 as it additionally samples the different local atomic environments at the transition states of the diffusion paths.
Tests 7 and 8 on defect formation show that the ISF can be captured by most models while the SIA requires in most cases a high weight on bulk structures.
Only two models near RMSE(bulk)=0.005eV and RMSE(elastic)=0.001eV pass all tests. The model with smaller RMSE(bulk) is selected as final BOP for Re and assessed in the following. The optimized parameters of the final BOP model are given in Tab~.\ref{tab:final_parameters}. 

\begin{table}[h!]
\begin{tabular}{c c c c c c c}
\hline\hline
\\
$H(R)$ & $c_0$ & $\lambda_0$ & $n_0$ & $c_1$ & $\lambda_1$ & $n_1$ \\ \hline
$dd\sigma$ & -25.6844 & 1.2112 & 0.9128 & -0.0545 & 0.0022 & 5.8615 \\ 
$dd\pi$ & 45.9185 & 1.5209 & 1.0781 & 1.3012 & 0.1368 & 2.8633 \\
$dd\delta$ & -11.8617 & 1.5604 & 0.9132 & -9.7241 & 1.2671 & 1.7076 \\
\hline
$E_\mathrm{pair}$ & $c_\mathrm{rep}$ & $\lambda_\mathrm{rep}$ & $n_\mathrm{rep}$ & & & \\
\hline
& 65538.09 & 4.7264 & 0.8626 & & & \\
\hline
$E_\mathrm{emb}$ & $a_\mathrm{emb}$ & $b_\mathrm{emb}$ & & & & \\
\hline
& 2.4338 & 0.1382 & & & & \\
\hline\hline
\end{tabular}
\caption{\label{tab:final_parameters} Parameters of final BOP for Re used in the assessment in Sec.~\ref{sec:finalBOP}. The corresponding BOP functional form is given in Secs.~\ref{sec:methodology} and \ref{sec:protocol}.}
\end{table}

\subsection{\label{sec:finalBOP}Assessment of final BOP}

In order to assess the local accuracy and the global transferability of the selected final BOP, an analysis is performed for the elastic constants and phonon spectrum of the hcp-Re ground-state, for point defects and stacking faults, as well as for the structural stability of TCP phases and random structures, and the BOP predictions are compared to DFT.
The lattice parameters of the hcp ground state are very well captured by all potentials as shown in Tab.~\ref{tab:propteries}. The elastic constants predicted by BOP are overall in good agreement with DFT with the largest deviations for $C_{12}$ and $C_{33}$.
\begin{table}[ht]
\begin{tabular}{c c c}
\hline\hline
& \,\,  DFT \, \, & \,\,  BOP \\ \hline
$a$ (\AA) & 2.782 & 2.786 \\
$c/a$ & 1.617 & 1.608 \\ 
$C_{11}$ (GPa)& 625 & 627 \\
$C_{12}$ (GPa)& 232 & 303 \\
$C_{13}$ (GPa)& 213 & 240 \\
$C_{33}$ (GPa)& 677 & 592 \\
$C_{44}$ (GPa)& 170 & 142 \\
$C_{66}$ (GPa)& 196 & 162 \\ 
$C_{12}-C_{66}$ (GPa)& 36 & 141 \\
$C_{13}-C_{44}$ (GPa)& 43 & 98 \\
$B$ (GPa)& 364 & 361 \\
\hline\hline
\end{tabular}
\caption{\label{tab:propteries} Comparison of lattice parameters and elastic constants of hcp ground state between DFT, BOP and EAM.}
\end{table}

\begin{figure}
\includegraphics[width=\columnwidth]{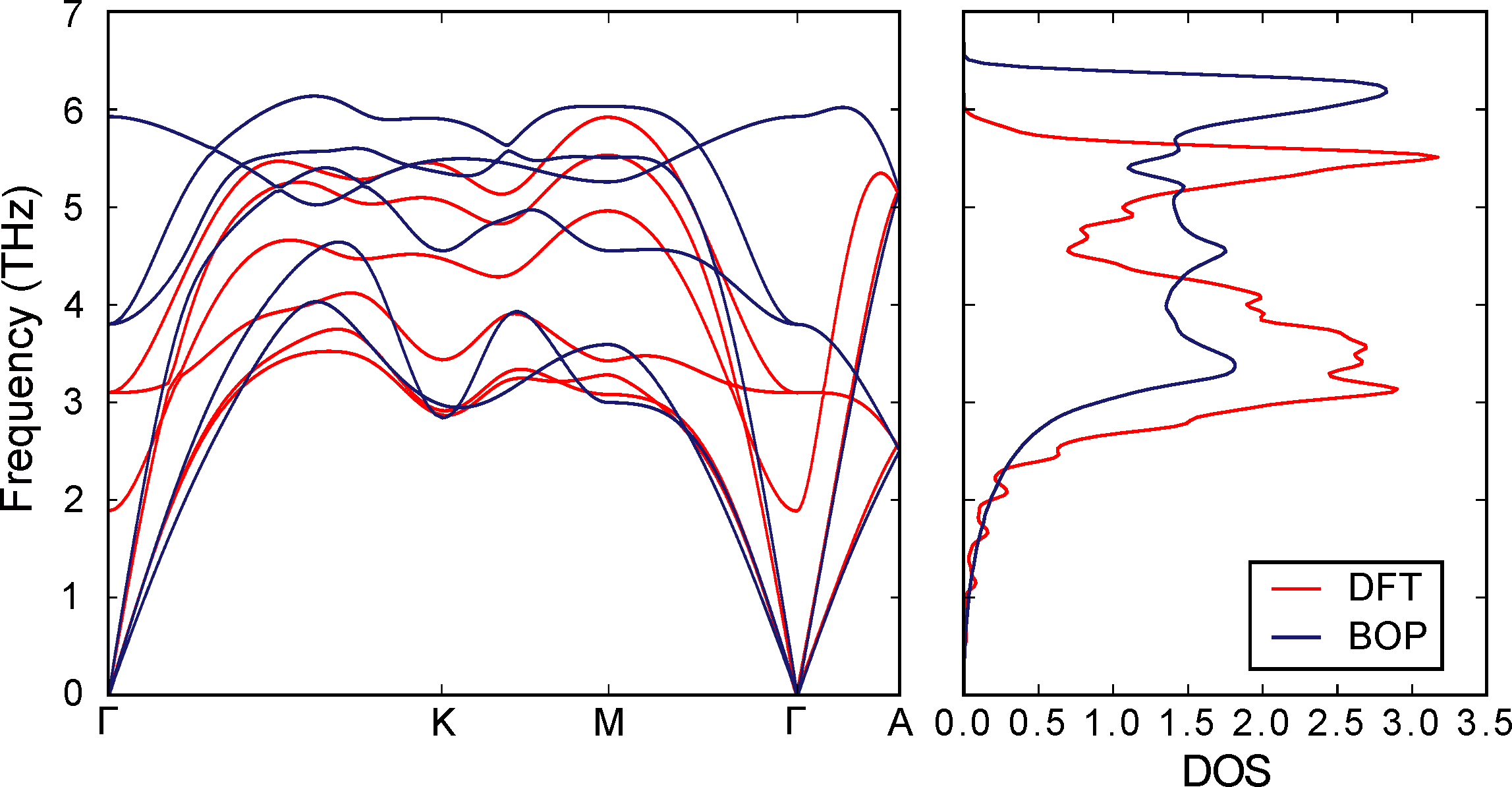}
\caption{Phonon spectrum and phonon density of states obtained by DFT and BOP.}
\label{fig:phonons}
\end{figure}

As one of the indicators of the performance in finite-temperature simulations we compute the phonon spectrum of Re-hcp, see Fig.~\ref{fig:phonons}, that was not included in the optimization procedure. The phonon density-of-states predicted by the final BOP is in good agreement with similar width but a small shift to higher frequencies. The phonon branches are overall in good qualitative agreement aside from the shift to higher frequencies.

The formation energies of points defects and stacking faults compiled in Tab.~\ref{tab:defects} sample the transferability to local atomic environments that were not included in the reference data for optimizing the potential. The BOP correctly predicts the vacancy as lowest-energy point defect. The energy barrier for vacancy diffusion in the basal plane is spot on while the perpendicular path is less favorable in BOP. The energetic ordering of SIA configurations is reproduced by the BOP except for the highest-energy basal tetrahedral. The absolute values of the formation energies are consistently overestimated by the BOP which we expect to be improved by adding more reference data with short interatomic distances. The formation energies of the stacking faults are in the correct order of magnitude and slightly underestimated by BOP. Both EAM potentials show larger deviations from the DFT reference data and several cases of SIA configurations that were not metastable in the relaxation by transferred to another SIA configuration as indicated in Tab.~\ref{tab:defects}.
\begin{table}
\begin{tabular}{c c c}
\hline
\\
& DFT & BOP \\ 
\hline
point defects (eV) \\
\hline
vacancy & 3.22 & 3.91 \\
vacancy diffusion (basal plane) & 2.02 & 2.02 \\
vacancy diffusion (perpendicular) & 1.71 & 2.17 \\
tetrahedral SIA & 6.76 & 8.93 \\
split dumbbell SIA & 6.78 & 8.96 \\
octahedral SIA & 8.16 & 9.46 \\
basal split dumbbell SIA & 9.41 & 10.97 \\
basal tetrahedral SIA (BT) & 10.17 & 10.44 \\
\hline
stacking faults (mJ/m$^\mathrm{2}$) \\
\hline
intrinsic & 55 & 21 \\ 
extrinsic & 349 & 278 \\ 
\hline
\end{tabular}
\caption{\label{tab:defects} Comparison of planar and point defects for the ground state hcp structure between DFT and BOP.}
\end{table}

\begin{figure}[ht]
\begin{subfigure}[b]{0.45\columnwidth}
\includegraphics[width=\textwidth]{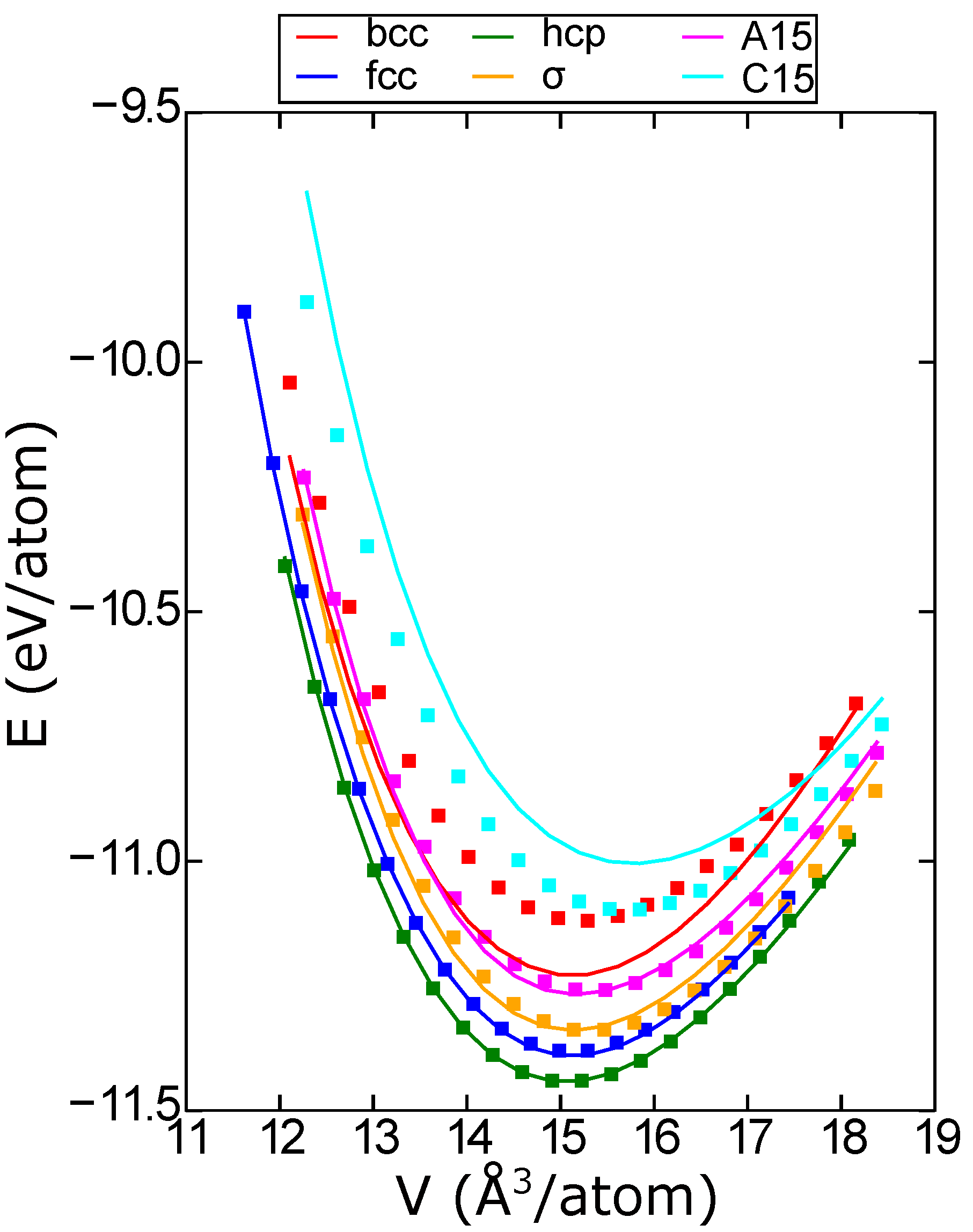}
\caption{a) Fit set}
\label{fig:bulk_fitset}
\end{subfigure}
\begin{subfigure}[b]{0.45\columnwidth}
\includegraphics[width=\textwidth]{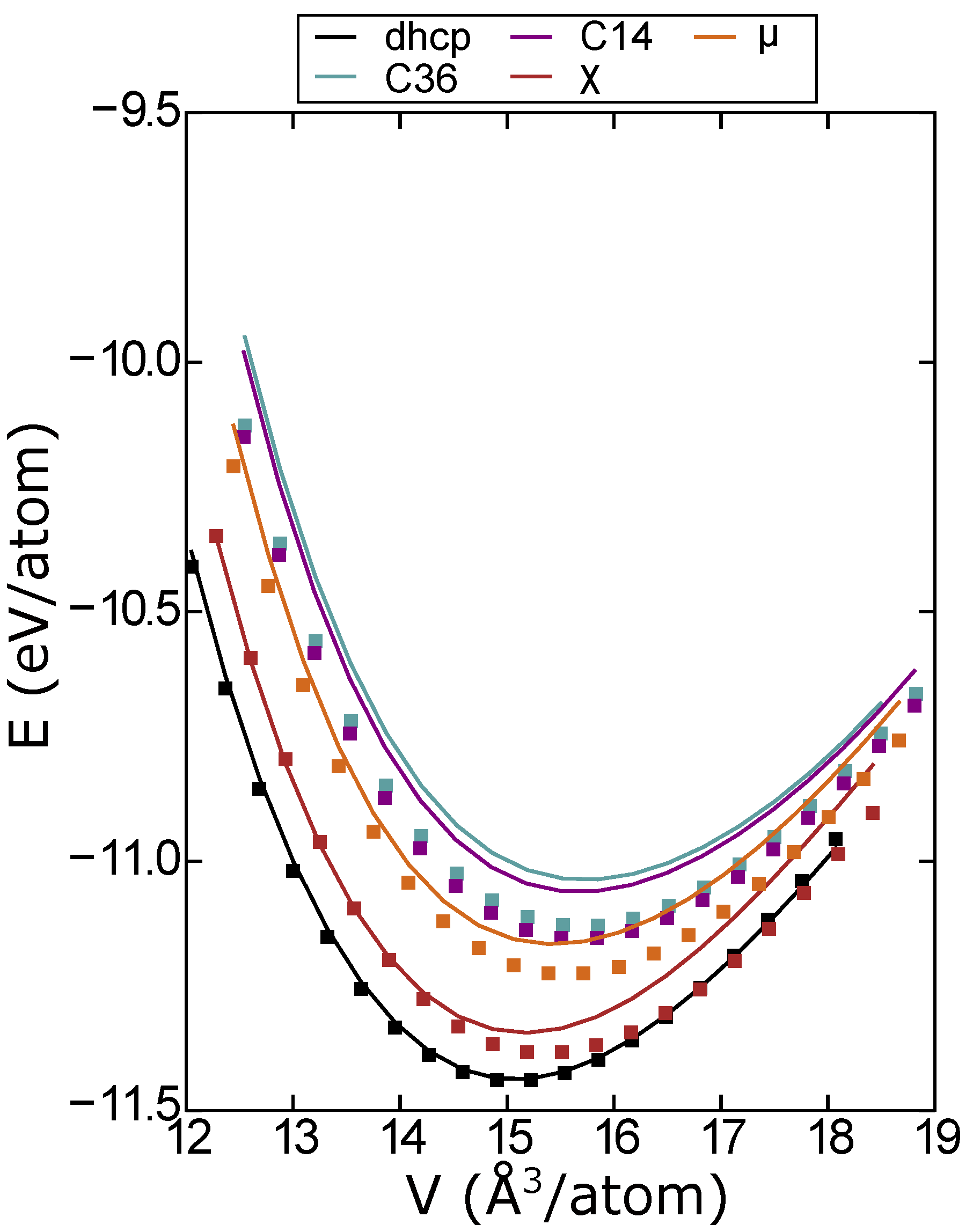}
\caption{b) Test set}
\label{fig:bulk_testset}
\end{subfigure}
\caption{Energy-volume curves of structures (a) in the fit set and (b) in the test set. DFT reference data is shown as symbols.}
\label{fig:structutal_stability}
\end{figure}

The transferability to other bulk structures is quantified by comparisons for structures that are not in the reference data, particularly dhcp, TCP phases (C14, C36, $\mu$, $\chi$) and the random structures with 1-atom unit cells (Fig.~\ref{fig:initial_model_performance}). The energy-volume curves of the structures in the reference data are reproduced very well (Fig.~\ref{fig:bulk_fitset}) with bcc as ground state and the correct ordering of all other structures. The larger deviations for the higher-energy structures A15 and C15 are a consequence of the energy-based weighting of reference-structures with Eq.~\ref{eq:weighting}. The transferability of the BOP becomes apparent in the energy-volume curves of the structures that were not included in the reference data in Fig~\ref{fig:bulk_testset}. The dhcp structure is reproduced with high accuracy and all other structures with good accuracy and correct energetic ordering. The good transferability from the fitted TCP phases (A15, C15) to the tested TCP phases (C14, C36, $\mu$, $\chi$) can partly be attributed to the similarity of the local coordination polyhedra in this class of crystal structures.

\begin{figure}[ht]
\includegraphics[width=\columnwidth]{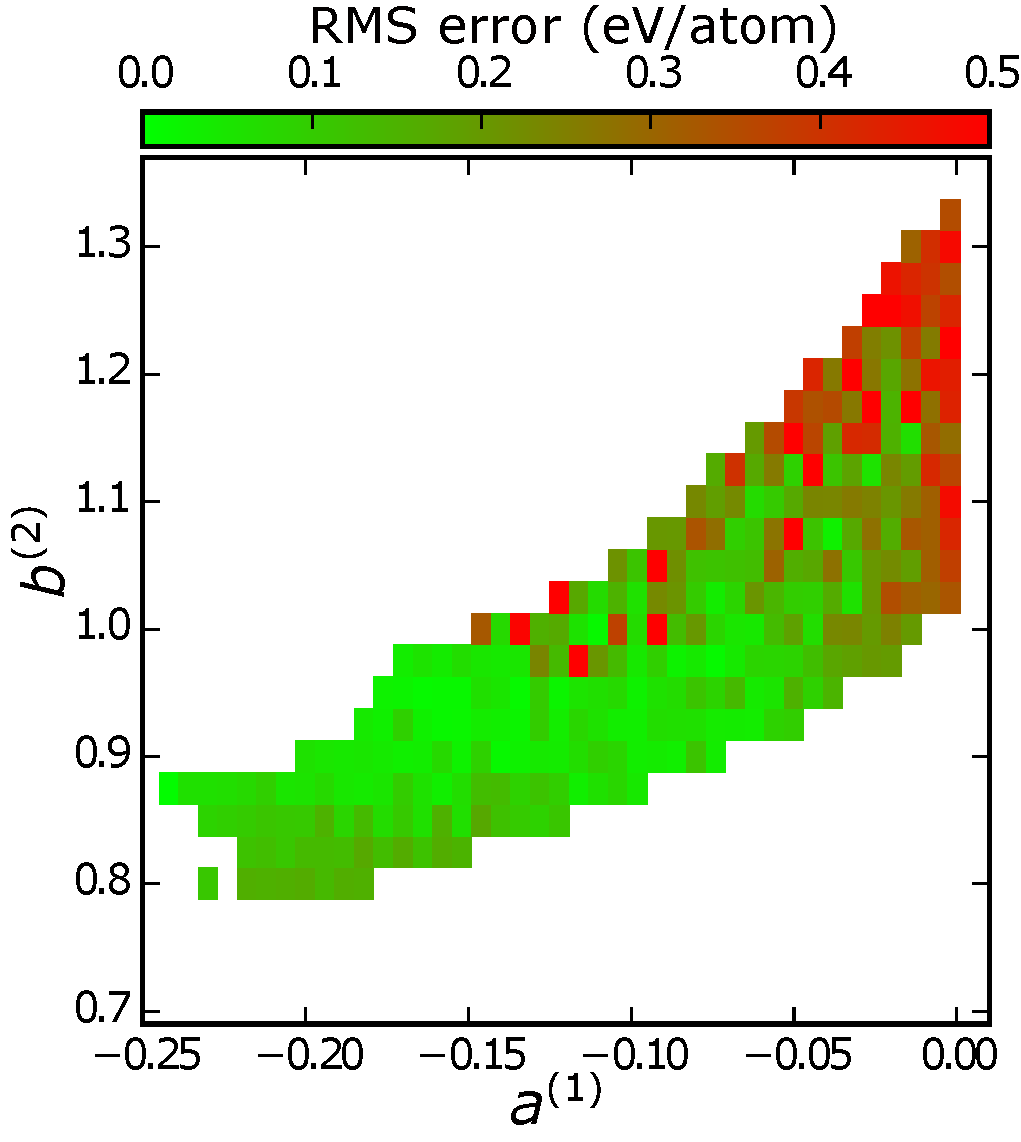}
\caption{Transferability of final model to random structures.}
\label{fig:random_final_model}
\end{figure}

The transferability of the BOP across the entire phase space of 1-atom unit cells introduced in Sec.~\ref{sec:randomfit} is shown in Fig.~\ref{fig:random_final_model}. The transferability to close-packed structures is very good while the RMS error is considerably larger for open structures that are energetically less favorable for Re. Comparing the RMS error of the final BOP to the RMS error of the basic BOP (Fig.~\ref{fig:rms_transferability_map_natoms_3}), we find a similar range of RMS error but a different distribution across the phase space of 1-atom unit cells.
The intrinsic transferability of the basic BOP that could be further improved with the first refinement strategy is apparently compromised in the second refinement strategy. The reason is the bias of the second strategy to the local atomic environments of elastic deformations and TCP phases. These are located at the lowest values of the phase space of 1-atom unit cells (e.g. hcp at $a^{(1)}=-0.24$) or outside at even lower values of $a^{(1)}<-0.25$ as shown in Ref.~\onlinecite{Jenke-18}. The emphasis on high-precision for structures in this region leads to the larger RMSE for the more open structures. The analysis with the RMSE in the map of local atomic environments highlights this difference between the two refinement strategies and provides transparent access to the balance of target properties. 

\section{\label{sec:conclusion}Conclusions}

A parameterization protocol for analytic bond-order potentials is presented that is closely related to the underlying coarse-grained description of the electronic structure. Starting with an initial $sd$-valent Hamiltonian obtained by DFT calculations, a pairwise repulsion is added to establish an initial binding-energy relation. The Hamiltonian is then simplified by replacing the contribution of the $s$ electrons by an isotropic embedding term. A basic BOP is then obtained by all parameters to energy-volume data of hcp, fcc and bcc. The good transferability of this basic BOP is demonstrated by a complete sampling of the phase space of 1-atom unit cells using a map of local atomic environments.

Different strategies of refining the basic BOP are presented and compared. It is demonstrated that the global transferability across the phase space of 1-atom unit cells can be systematically improved by simple homogeneous samplings with increasing density. An alternative strategy is presented of including elastic constants and further crystal structures in the optimization and shown to improve the local accuracy. The combination of the Pareto front for different weightings of the reference data with additional tests illustrates the balancing of target properties and leads to a final BOP for Re. The final BOP is shown to give robust predictions for elastic constants, phonons, point defects, stacking faults and the energetic ordering of various crystal structures. An analysis of the final BOP with the RMS error across the entire phase space of 1-atom unit cells highlights the compromise between local accuracy and  global transferability.

Details of the parameterization protocol are specific to BOP and Re but the overall concepts are generally applicable to the parameterization of interatomic potentials.

\section*{Acknowledgments}

We acknowledge financial support by the German Research Foundation (DFG) through research grant HA 6047/4-1 (project number 289654611) and project C1 of the collaborative research center SFB/TR 103 (project number 190389738).

\nocite{*}
\bibliographystyle{elsarticle-num}
\bibliography{bibliography}

\end{document}